\documentclass[%
reprint,
superscriptaddress,
nofootinbib,
 amsmath,amssymb,
 aps,
]{revtex4-2}
\usepackage{graphics,epsfig,psfrag}
\usepackage{color}
\usepackage{graphicx}
\usepackage{dcolumn}
\usepackage{bm}
\usepackage{hyperref}

\begin{document}

\title{Shape of higher-order images\\
of equatorial emission rings around a Schwarzschild black hole:\\
Analytical description with polar curves}

\author{Oleg Yu. Tsupko}
\email{tsupko@iki.rssi.ru; tsupkooleg@gmail.com}
\affiliation{Space Research Institute of Russian Academy of Sciences, Profsoyuznaya 84/32, Moscow 117997, Russia}

\date{\today}

\begin{abstract}
Higher-order photon rings can be expected to be detected in a more detailed image of the black hole found in future observations. These rings are lensed images of the luminous matter surrounding the black hole and are formed by photons that loop around it.
In this paper we have succeeded to derive an analytical expression for the shape of the higher-order rings in the form that is most convenient for application: the explicit equation of the curve in polar coordinates. The formula describes the apparent shape of the higher-order image of the circular ring with the given radius around Schwarzschild black hole as viewed by distant observer with an arbitrary inclination. For the derivation, the strong deflection limit of the gravitational deflection is used. Our formula is a simple and efficient alternative to the numerical calculation of ray trajectories, with the main application to studying the shape of $n=2$ and $n=3$ photon rings.
\end{abstract}

\maketitle


\section{Introduction}

One consequence of the gravitational deflection of light is the formation of multiple images of the same source. In the case when gravitational lensing occurs on a black hole, so-called higher-order images can appear. These images are formed by photons that have experienced one or more turns around the black hole before reaching the observer.

Higher-order images have been extensively studied, see, e.g., reviews \cite{Perlick-2004a, Bozza-2010, BK-Tsupko-Universe-2017}; see also Introduction in our previous article \cite{BK-Tsupko-2022}. Among early papers on the subject, one can refer to \cite{Darwin-1959, Atkinson-1965, MTW-1973, Ohanian-1987, Bao-Hadrava-1994a, Bao-Hadrava-1994b}. More active research started about two decades ago. Numerical studies of higher-order images were presented in the paper by Virbhadra and Ellis \cite{Virbhadra-2000} (the authors introduced the term 'relativistic images'), see also \cite{Virbhadra-2001, Virbhadra-2009}. In analytical studies, the strong deflection limit of gravitational deflection was used: an analytical formula describing the logarithmically diverging deflection angle of photons making one or more revolutions around the black hole \cite{Darwin-1959, Bozza-2001, Bozza-2002}. Analytical calculations of relativistic images were developed in the works of Bozza et al \cite{Bozza-2001, Bozza-2002, Bozza-2003, Bozza-Mancini-ApJ-2004, Bozza-Mancini-ApJ-2005, Bozza-2005, Bozza-2006, Bozza-Sereno-2006, Bozza-2008}. The strong deflection limit (also known as strong field limit) was also used in the series of paper by Eiroa et al \cite{Eiroa-2002, Eiroa-2004, Eiroa-2005, Eiroa-2006}. Gravitational lensing beyond the weak deflection approximation was studied also in Frittelli et al \cite{Frittelli-2000} and Perlick \cite{Perlick-2004a, Perlick-2004b}. Further, higher-order images have been also investigated both numerically and analytically by many groups 
\cite{Amore-2007, Iyer-Petters-2007, Gyulchev-2007, BK-Tsupko-2008, Chen-Jing-2009, Ghosh-2010, Bin-Nun-2011, Ding-Kang-2011, Wei-2012, Tsupko-BK-2013, Sadeghi-2014, Alhamzawi-2016, Tsukamoto-2016, Chakraborty-2017, Barlow-2017, Bozza-2017, Uniyal-2018, Bergliaffa-2020, Tsukamoto-2021a, Tsukamoto-2021b, Islam-2021, Furtado-2021, Aratore-Bozza-2021, Tsukamoto-2022}. The strong deflection limit for massive particles was first found by Tsupko \cite{Tsupko-2014}, see also \cite{Liu-2016, Crisnejo-Gallo-Jusufi-2019, Wang-Liu-2020}.

Most of the articles considered the case when the source is compact and is at a large distance from the black hole, where the gravity of the black hole can be neglected. The important generalization has been made in the work of Bozza and Scarpetta \cite{Bozza-Scarpetta-2007} who have developed the strong deflection limit for sources at arbitrary distance from the black hole. It was further used in the papers of Bozza et al \cite{Bozza-2010, Aldi-Bozza-2017, Aratore-Bozza-2021}. In particular, this approach was used for analytical investigations of higher-order images of the accretion disk in the work of Aldi and Bozza \cite{Aldi-Bozza-2017}.

\begin{figure}
\begin{center}
\includegraphics[width=0.46\textwidth]{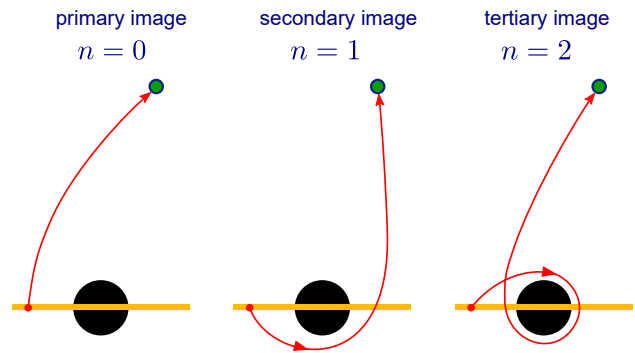}
\end{center}
\caption{The first three images in an infinite series of images of the accretion disk around the black hole, numbered by the number $n$ of half-orbits. Direct (primary) image has $n=0$. Secondary image has $n=1$. Images with $n \ge 2$ are denoted as higher-order rings, starting from $n=2$ photon ring (tertiary image).}
\label{fig:rings}
\end{figure}

The recent emergence of great interest in the subject of higher-order images is associated with an observational breakthrough in black hole imaging. In 2019, the Event Horizon Telescope has been presented the shadow of supermassive black hole in the M87 galaxy \cite{Falcke-2000, EHT-1, EHT-2, EHT-3, EHT-4, EHT-5, EHT-6, Psaltis-EHT-2020, Kocherlakota-2021} (see also recent results about Sgr A* \cite{EHT-2022-1, EHT-2022-2, EHT-2022-3, EHT-2022-4, EHT-2022-5, EHT-2022-6}). After that, attention has been attracted to the study of higher-order photon rings which can be expected in a more detailed image, being concentrated near the edge of the black hole shadow \cite{Gralla-2019, Johnson-2020, Gralla-Lupsasca-2020, Gralla-Lupsasca-Marrone-2020, Pesce-2021, Wielgus-2021, Gan-Wang-2021, Guerrero-2021, Frion-2021, Wang-Chen-2022, Bogush-2022, Chen-Roy-2022, Hou-Liu-2022, Walia-2022, Andrianov-2022, Broderick-2022, Paugnat-Lupsasca-2022, Vincent-Gralla-2022, Ayzenberg-2022}. These rings are lensed images of the luminous matter surrounding the black hole. It is now quite common to denote these images by the number of half-orbits $n$, e.g., \cite{Johnson-2020, Gralla-Lupsasca-2020, Gralla-Lupsasca-Marrone-2020, Pesce-2021, Wielgus-2021, Broderick-2022, Paugnat-Lupsasca-2022, Vincent-Gralla-2022, Ayzenberg-2022}. The direct image is denoted as $n=0$, secondary image has $n=1$, and we refer all images $n \ge 2$ to as higher-order photon rings (Fig.\ref{fig:rings}).

Numerical simulations are used to obtain a detailed image of the black hole that can be compared with observations, see Event Horizon Telescope papers \cite{EHT-1, EHT-2, EHT-3, EHT-4, EHT-5, EHT-6, Kocherlakota-2021, EHT-2022-1, EHT-2022-2, EHT-2022-3, EHT-2022-4, EHT-2022-5, EHT-2022-6} and, e.g., \cite{Falcke-2000, Broderick-Loeb-2005, Moscibrodzka-2009, Dexter-2009, Broderick-Johannsen-2014, James-2015, Narayan-2019, Dokuchaev-2020, Bronzwaer-Falcke-2021, Andrianov-2021, Vincent-2021, Bronzwaer-Davelaar-2021}. At the same time, a great number of works are devoted to analytical studies of the black hole shadow, in which the size and shape of the shadow boundary (also known as the critical curve) are studied. E.g., the deformed shape of the shadow of the Kerr black hole has known analytical representation, e.g., \cite{Bardeen-1973, Chandra-1983, Cunha-Herdeiro-2018, Perlick-Tsupko-2022}; for an observer at an arbitrary distance from the black hole, see \cite{Gren-Perlick-2014, Gren-Perlick-2015, Perlick-Tsupko-2022}. Analytical investigations of shape of Kerr black hole shadow can be found, e.g., in \cite{Bardeen-1973, Young-1976, Chandra-1983, Dymnikova-1986, Vries-2000, Takahashi-2004, Zakharov-Paolis-2005-New-Astronomy, Bozza-2006, Hioki-Maeda-2009, Johannsen-2010, Frolov-Zelnikov-2011, Johannsen-2013, Gren-Perlick-2014, Li-Bambi-2014, Gren-Perlick-2015, Tsupko-2017, Gralla-Lups-2018, Cunha-Herdeiro-2018, Wei-2019-Rapid, Farah-2020, Gralla-Lups-2020c, Perlick-Tsupko-2022}. For another examples of theoretical consideration of black hole shadow see, e.g., \cite{Perlick-Tsupko-BK-2015, Cunha-2015, Abdujabbarov-2015, Shipley-Dolan-2016, Perlick-Tsupko-2017, Yan-2018, Mars-2018, Yunes-2018, Cunha-2018, Perlick-Tsupko-BK-2018, BK-Tsupko-2018, Abdikamalov-2019, Tsupko-Fan-BK-2020, Qi-Zhang-2020, Vagnozzi-2020, Neves-2020, Li-Guo-2020, Chang-Zhu-2020, Tsupko-BK-2020-IJMPD, Kumar-Ghosh-2020, Cunha-2020, Lima-2021, Wang-2021, Eichhorn-2021, Ozel-2021, Devi-2021, Pantig-2021, Anacleto-2021, Khodadi-2021, Tsupko-2021, Cardoso-2021, Roy-2021, Ghosh-Sarkar-2021, Uniyal-Pantig-2022, Vagnozzi-review-2022}. We refer to the recent review by Perlick and Tsupko \cite{Perlick-Tsupko-2022} as an overview of analytical studies of the shadow, see also the review by Cunha and Herdeiro \cite{Cunha-Herdeiro-2018}.

The shape of the higher-order photon rings is very close to the boundary of the black hole shadow (critical curve), however, it differs from it and is of considerable interest from the observational point of view, e.g. \cite{Johnson-2020, Gralla-Lupsasca-Marrone-2020}. Even in the simplest case of Schwarzschild black hole, for which the shadow boundary is a circle, the photon ring is a circle only for an observer on the axis of symmetry. For observer with inclination, this is a deformed curve, the shape of which would be very useful to know in an analytical form.

In this paper, we will restrict our attention to equatorial rings of emission and will use the term "photon ring" to refer to the higher-order images of such rings.

Here, we derive the analytical expression for the shape of higher-order rings in the form that is most convenient for application: the explicit equation of the curve in polar coordinates. The formula describes the apparent shape of higher-order image of circular orbit with given radius around the Schwarzschild black hole for distant observer with an arbitrary inclination. A simple analytical formula for the curve allows us to find a number of properties of the ring images.
\footnote{Throughout the article, by 'circular orbit' or 'circular ring' we mean thin circular radiating ring of given constant radius orbiting the black hole in the equatorial plane (e.g., Luminet \cite{Luminet-1979}). Specifically, in the title, we refer to this ring as the "equatorial emission ring". This ring should be not confused with the 'photon ring', which is lensed higher-order \textit{image} of circular ring of emission on the observer's sky. See also the discussion on p.7 of \cite{Perlick-Tsupko-2022}.}

In our derivation, we use the strong deflection limit of light deflection for arbitrary source position mentioned above \cite{Bozza-Scarpetta-2007, Bozza-2010}. Since photon rings are formed by light sources near the black hole, this method is very convenient to use for analytical calculation of their properties (strong deflection approximation is appropriate for $n \ge 2$ rings). E.g., in our previous work \cite{BK-Tsupko-2022}, we have considered the thin accretion disk around Schwarzschild black hole and the observer located on the symmetry axis (polar view). For that configuration, we calculated analytically the angular radii, thicknesses, and solid angles of higher-order rings in the form of compact analytical expressions; we also made estimates for fluxes. In our present article, we consider the observer with arbitrary inclination angle and focus only on the deformed shape of photon ring images of higher orders ($n \ge 2$).

The paper is organized as follows. In the next Section we derive an explicit analytic formula for the shape of the higher-order rings. In the Section \ref{sec:discussion} we present graphs, and discuss properties of higher-order rings that are derived analytically using our results. The Section \ref{sec:conslusion} is our Conclusions. In the Appendix, we present the alternative derivation of our formula based on results of Aldi and Bozza \cite{Aldi-Bozza-2017}, finding complete agreement.

\begin{figure*}
\begin{center}
\includegraphics[width=0.90\textwidth]{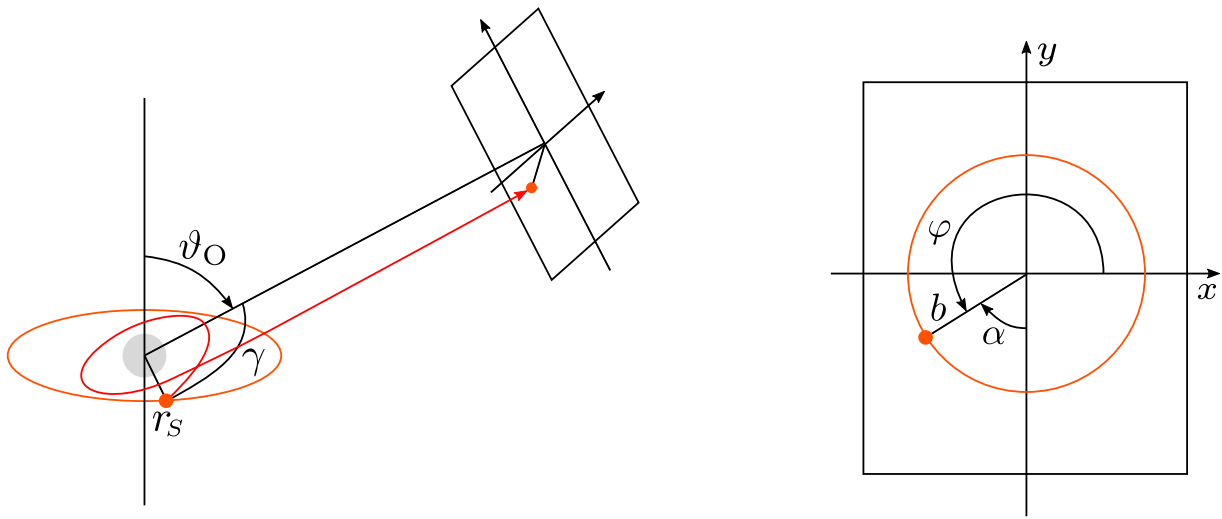}
\end{center}
\caption{Geometry of the problem and the variables used. The left panel shows the circular orbit of radius $r_S$ and the light ray from an element of circular orbit making one complete revolution around the black hole and arriving at the observer with inclination $\vartheta_\mathrm{O}$. Such ray gives the tertiary ($n=2$) image of the source. The angle $\gamma$ is measured in the plane of the ray trajectory. The picture on the right shows the observer's reference frame. The angle $\gamma$ in the plane of the ray is related to the angle $\alpha$ in the observer's sky by the formula (\ref{gamma-alpha}), see article \cite{Luminet-1979} and Fig.3 there. The angle $\alpha$ is then converted to the more familiar angle $\varphi$, measured counterclockwise from the horizontal axis, see eq.(\ref{alpha-to-varphi}). The final shape $b_2(\varphi)$ of the $n=2$ lensed image of the circular orbit is given by the formula (\ref{b-varphi}).}
\label{fig:geometry}
\end{figure*}

\section{Polar curve for the shape of higher-order photon rings}

In this Section we derive the explicit analytical expression for the shape of the higher-order photon rings, using the strong deflection limit of gravitational deflection. First we consider $n=2$ ring and make the full derivation for it, then we consider $n=3$ ring, then we obtain the general formula for the arbitrary $n$.

We write the Schwarzschild metric as
\begin{equation}  \label{schw-metric}
ds^2 = - \left(1 - \frac{2m}{r} \right) \, c^2 dt^2 + \frac{dr^2}{1-2m/r} \, +
\end{equation}
\[
+ \, r^2
\left( d \vartheta^2 + \sin^2 \vartheta \, d \phi^2 \right),   \;   m=\frac{GM}{c^2}   \,   ,
\]
where $m$ is a mass parameter of dimension of length, and $M$ is the black hole mass. In these notations \cite{BK-Tsupko-2022}: the horizon is located at radial coordinate $2m$; unstable circular orbit of photons has $r_{\mathrm{ph}}=3m$ (photon sphere radius); innermost stable circular orbit for massive particles is located at $r_\mathrm{ISCO} = 6m$. The critical value of the impact parameter, corresponding to photons flying from infinity and arriving at the photon sphere, is $b_\mathrm{cr} = 3\sqrt{3}m$. This value defines the linear size of the shadow of Schwarzschild black hole, e.g.\cite{Perlick-Tsupko-2022}.

We consider the circular ring of fixed constant radius $r_S$ in the equatorial plane $\vartheta=\pi/2$ (Fig.\ref{fig:geometry}). The observer is assumed to be located at a large distance $D_d$ from the black hole, $D_d \gg m$. The observer's inclination angle is $\vartheta_\mathrm{O}$. To a distant observer, the circle will be visible as the sequence of lensed images. Primary and secondary images of circular orbits can be found, e.g., in Luminet's paper \cite{Luminet-1979} who call them 'isoradial curves', see also, e.g., \cite{Cunningham-1973, Bao-Hadrava-1994a}. In the case of polar observer, $\vartheta_\mathrm{O}=0$, circular orbits are seen as non-deformed circles; the analytical results for this case are presented in \cite{BK-Tsupko-2022}.

We are interested in the deformed shape of higher-order ring image appeared to the distant observer. Observed shape in observer's frame is described by the impact parameter $b$ (measured in units of $m$) and the polar angle $\varphi$ (Fig.\ref{fig:geometry}). If one is interested in angular values, then the following relation (valid for observer at large distances) should be applied: $\theta=b/D_d$, where $\theta$ is the observed angular radius. For example, the angular radius of black hole shadow for the observer at large distances is $\alpha_{\mathrm{sh}} = 3\sqrt{3}m/D_d$.\\

Our derivation consists of three steps. In the first step, we will obtain the impact parameter $b$ as a function of the angle $\gamma$ in plane of ray path.

Consider a point source with radial coordinate $r_S$ located outside the photon sphere. Since we are interested in higher-order images, we consider a light ray that starts at $r_S$, first moves towards the black hole, makes one or more revolutions around it near the photon sphere, and then flies away to infinity. The distance of the closest approach $R$ (the minimum value of the radial coordinate) of such ray is close to the photon sphere radius $r_\mathrm{ph}$, while the impact parameter $b$ is close to its critical value $b_{cr}$.

Working in the plane of ray trajectory, we write the change of the angular coordinate $\Delta \tilde{\phi}$ (azimuthal shift) of such ray in the strong deflection limit as
\cite{Bozza-Scarpetta-2007, Bozza-2010, Aratore-Bozza-2021}:
\begin{equation} \label{bozza-2007-03}
\Delta \tilde{\phi} = - \ln \epsilon
+ \ln f(r_S) \, ,
\end{equation}
where
\begin{equation} \label{bozza-2007-04}
\epsilon = \frac{b - b_{cr}}{b_{cr}} \ll 1 \, , \quad b_{cr} = 3\sqrt{3}m \, ,
\end{equation}
\begin{equation} \label{f-rS}
f(r_S) =  \frac{6^5 \left(1 - \frac{3m}{r_S} \right) }{
\left(3+\sqrt{3} \right)^2 
} \left( 3+ \sqrt{3 + \frac{18m}{r_S}} \right)^{-2} \, .
\end{equation}
Note that the variable $\tilde{\phi}$ is defined in the ray plane and therefore should not be confused with the variable $\phi$ in (\ref{schw-metric}).

We start from Eq.(\ref{bozza-2007-03}) and find
\begin{equation} \label{b-general}
b = b_{cr} \left[ 1 + f(r_S) \,  e^{-\Delta \tilde{\phi}}   \right] \, .
\end{equation}

The Fig.\ref{fig:geometry} shows the light ray which forms $n=2$ image. For this ray we have:
\begin{equation} \label{shift-01}
\Delta \tilde{\phi} = 2 \pi + \gamma \, .
\end{equation}
Using Eq.(\ref{shift-01}) in Eq.(\ref{b-general}), we find:
\begin{equation} \label{b-gamma}
b(\gamma) = 3\sqrt{3}m \left[ 1 + f(r_S) \,  e^{-2\pi - \gamma}   \right] \, .
\end{equation}

The second step of derivation is to relate the angle $\gamma$ characterizing the trajectory with some angle in the observer's frame. A convenient way to do this is to follow Luminet's paper \cite{Luminet-1979} who used the polar angle $\alpha$ and got the following simple relation:
\begin{equation} \label{gamma-alpha}
\cos \gamma = \frac{\cos \alpha}{\sqrt{\cos^2 \alpha + \cot^2 \vartheta_\mathrm{O}} }     \, ,
\end{equation}
see Fig.\ref{fig:geometry} here and Fig.3 in \cite{Luminet-1979}.

\begin{widetext}

Using eq.(\ref{gamma-alpha}) in (\ref{b-gamma}), we find the apparent shape of image of circular orbit as the function $b(\alpha)$:
\begin{equation} \label{b-alpha}
b(\alpha) = 3\sqrt{3}m \left\{ 1 + f(r_S) \,  \exp \left[ -2\pi - \arccos \left( \frac{\cos \alpha}{\sqrt{\cos^2 \alpha + \cot^2 \vartheta_\mathrm{O}} } \right) \right]   \right\} \,  .
\end{equation}

Our third step, for purposes of convenience mainly, is to rewrite the formula (\ref{b-alpha}) through the polar angle $\varphi$, measured in the 'usual' way: counterclockwise from the horizontal axis (Fig.\ref{fig:geometry}). We write:
\begin{equation} \label{alpha-to-varphi}
\cos \alpha = \cos(3 \pi/2 - \varphi) = - \sin \varphi \, .
\end{equation}
Using the relation $\arccos(-x)=\pi - \arccos{x}$, we obtain finally the shape of tertiary image ($n=2$ photon ring) of circular orbit with radius $r_S$ as the polar curve $b_2(\varphi)$:
\begin{equation} \label{b-varphi}
b_2(\varphi) = 3\sqrt{3}m \left\{ 1 + f(r_S) \,  \exp \left[ -3\pi + \arccos \left( \frac{\sin \varphi}{\sqrt{\sin^2 \varphi + \cot^2 \vartheta_\mathrm{O}} } \right) \right]   \right\} \, .
\end{equation}
To obtain a complete curve, the angle $\varphi$ must vary from $0$ to $2\pi$.\\

The next image ($n=3$ photon ring) is formed by photons that have passed behind the black hole and then turned around once more before reaching the observer. Such ray experiences the following change of azimuthal coordinate:
\begin{equation}
\Delta \tilde{\phi} = 3 \pi + (\pi - \gamma)  \, .
\end{equation}
Compared to $n=2$ image, the light ray forming $n=3$ image reaches the point in opposite part of the observer's sky. Therefore we have to replace $\alpha$ with $\pi + \alpha$ in the Eq.(\ref{gamma-alpha}). We get:
\begin{equation} \label{gamma-alpha-minus}
\cos \gamma = \frac{- \cos \alpha}{\sqrt{\cos^2 \alpha + \cot^2 \vartheta_\mathrm{O}} }     \, .
\end{equation}
Correspondingly, we find the dependence $b(\alpha)$ for $n=3$ image as
\begin{equation} \label{b-alpha-2}
b(\alpha) = 3\sqrt{3}m \left\{ 1 + f(r_S) \,  \exp \left[ -4\pi + \arccos \left( \frac{-\cos \alpha}{\sqrt{\cos^2 \alpha + \cot^2 \vartheta_\mathrm{O}} } \right) \right]   \right\} \, . 
\end{equation}
Transforming from $\alpha$ to $\varphi$ with eq.(\ref{alpha-to-varphi}), we obtain finally
the shape of $n=3$ photon ring as the polar curve $b_3(\varphi)$:
\begin{equation} \label{b-varphi-n3}
b_3(\varphi) = 3\sqrt{3}m \left\{ 1 + f(r_S) \,  \exp \left[ -4\pi + \arccos \left( \frac{\sin \varphi}{\sqrt{\sin^2 \varphi + \cot^2 \vartheta_\mathrm{O}} } \right) \right]   \right\} \, . 
\end{equation}


To obtain the formula for general $n$, we notice that for rings with $n=2, 4, 6, ...$ we have:
\begin{equation}
\Delta \tilde{\phi} = n \pi + \gamma \, ,
\end{equation}
and eq.(\ref{gamma-alpha}) should be used, whereas for $n=3, 5, 7, ...$ we have:
\begin{equation}
\Delta \tilde{\phi} = n \pi + (\pi - \gamma) = (n+1) \pi - \gamma \, ,
\end{equation}
and eq.(\ref{gamma-alpha-minus}) should be used.

As a result, we can write the general formula valid for all higher-order images of $n$-th order:
\begin{equation} \label{b-varphi-general-n}
b_n(\varphi) = 3\sqrt{3}m \left\{ 1 + f(r_S) \,  \exp \left[ -(n+1)\pi + \arccos \left( \frac{\sin \varphi}{\sqrt{\sin^2 \varphi + \cot^2 \vartheta_\mathrm{O}} } \right) \right]   \right\} \, , \quad n \ge 2 \, .
\end{equation}
For the reader's convenience, we remind our notations here.
Eq.(\ref{b-varphi-general-n}) describes the apparent shape of higher-order image (photon ring of $n$-th order) of circular orbit with constant radius $r_S$ in the equatorial plane of Schwarzchild black hole. Angle $\vartheta_\mathrm{O}$ defines the observer's inclination angle measured from the axis of symmetry. Variable $b_n$ is the impact parameter in the observer's sky; polar angle $\varphi$ is set in the usual way: counterclockwise from the horizontal axis (Fig.\ref{fig:geometry}). Function $f(r_S)$ is given by (\ref{f-rS}). Mass parameter $m$ is defined in (\ref{schw-metric}), and the shadow boundary has the radius $b_{cr}= 3\sqrt{3}m$.\\

If one considers an accretion disk with given inner $r_S^\mathrm{in}$ and outer $r_S^\mathrm{out}$ radii, then the photon ring will also have the inner $b_n^\mathrm{in}(\varphi)$ and outer $b_n^\mathrm{out}(\varphi)$ boundaries, determined by the values of these radii. The thickness of $n$-photon ring as the function of $\varphi$ will be:
\begin{equation} \label{thickness-general}
\Delta b_n(\varphi) \equiv
b_n^\mathrm{out}(\varphi) - b_n^\mathrm{in}(\varphi) 
=
3\sqrt{3}m  \left[ f(r_S^\mathrm{out}) - f(r_S^\mathrm{in}) \right]   \exp \left[ -(n+1)\pi + \arccos \left( \frac{\sin \varphi}{\sqrt{\sin^2 \varphi + \cot^2 \vartheta_\mathrm{O}} } \right) \right]  \, ,
\end{equation}
and the area of $n$-photon ring will be:
\begin{equation} \label{area-general}
\Delta S_n = \frac{1}{2} \int \limits_0^{2\pi} (b_\mathrm{out}^2(\varphi) - b_\mathrm{in}^2(\varphi) ) \, d\varphi \simeq 3\sqrt{3}m  \int \limits_0^{2\pi}  \Delta b_n(\varphi)  \, d\varphi \, ,
\end{equation}
\begin{equation}
\mbox{or} \quad \Delta S_n = 6\sqrt{3}m \, \pi \, \langle \Delta b_n(\varphi) \rangle  \, , \quad \mbox{where} \quad \langle \Delta b_n(\varphi) \rangle = \frac{1}{2\pi} \int \limits_0^{2\pi}  \Delta b_n(\varphi)  \, d\varphi \, .
\end{equation}

We note that an analogue of Eq.(\ref{b-varphi-general-n}) has been derived for the Kerr case in Appendix A of Hadar et al \cite{Hadar-Johnson-2021}. See in particular Eq.(A.14), which is not derived using the strong deflection limit but rather the matched asymptotic expansion derived in Appendix B of Gralla and Lupsasca \cite{Gralla-Lupsasca-2020}. Since this approximation breaks down for large source radius $r_S$, it would be interesting to apply a different method (such as the strong deflection limit) to obtain an analytic formula valid for all source radii.\footnote{We thank the anonymous Referee for pointing us to this result.}

\end{widetext}

\begin{figure*}
\begin{center}
\includegraphics[width=0.95\textwidth]{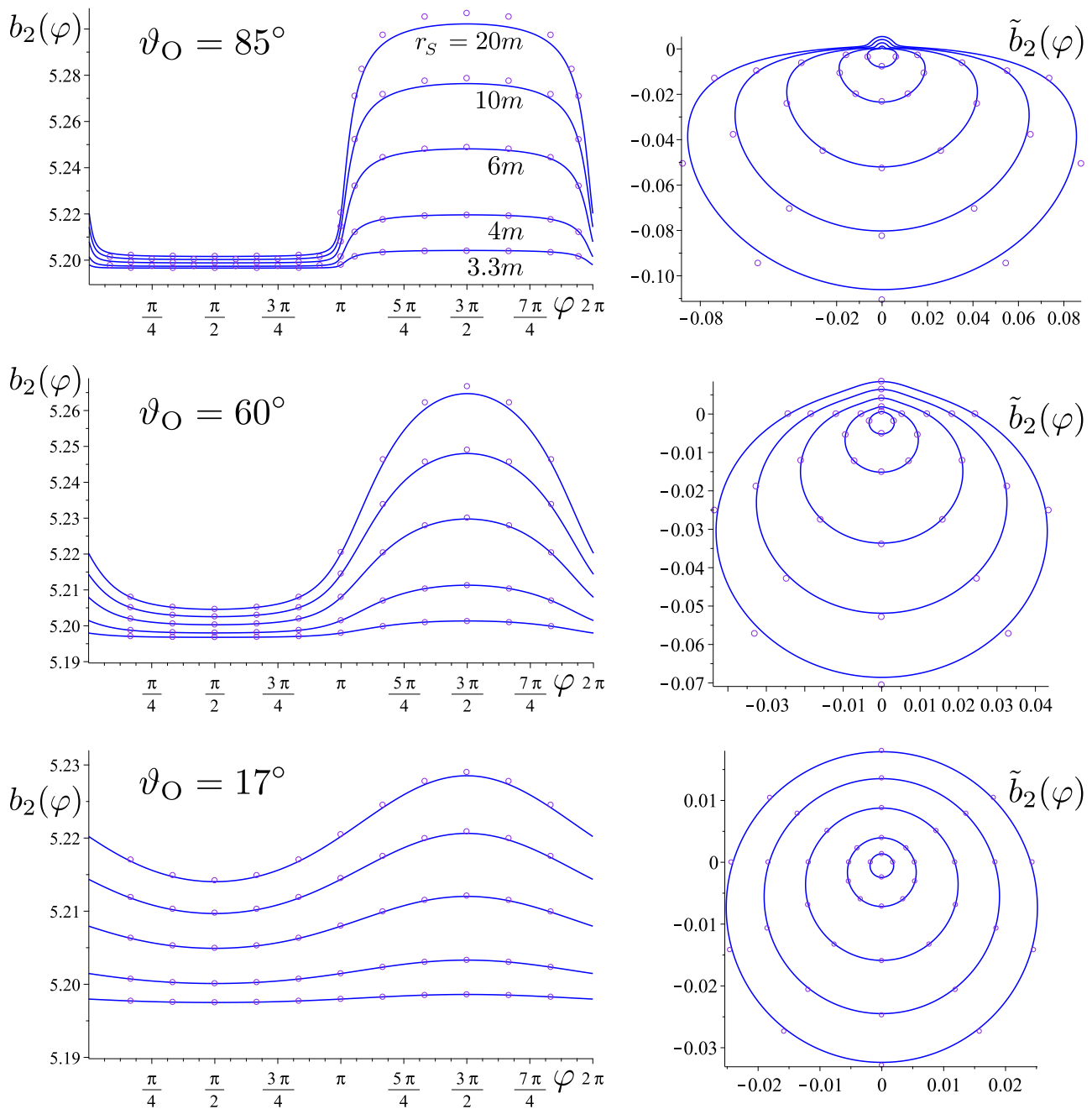}
\end{center}
\caption{Apparent shape of tertiary image ($n=2$ photon ring) of concentric equatorial rings of different radii $r_S$ for different inclinations $\vartheta_\mathrm{O}$ of the observer. The curves are plotted according to the analytical formula (\ref{b-varphi}). Each left panel shows the impact parameter $b_2$ as the function of angle $\varphi$ in the observer's sky, for the following values of ring radii $r_S$: $3.3m$, $4m$, $6m$, $10m$, $20m$. Each right panel shows the corresponding excess $\tilde{b}_2(\varphi)$ plotted in polar coordinates, see eq.(\ref{excess}), which can provide better understanding of the deformed shape of the image. The small circles show the results of numerical calculations, see Subsection \ref{subsec-errors} for details.}
\label{fig:th-several}
\end{figure*}

\begin{figure*}
\begin{center}
\includegraphics[width=0.98\textwidth]{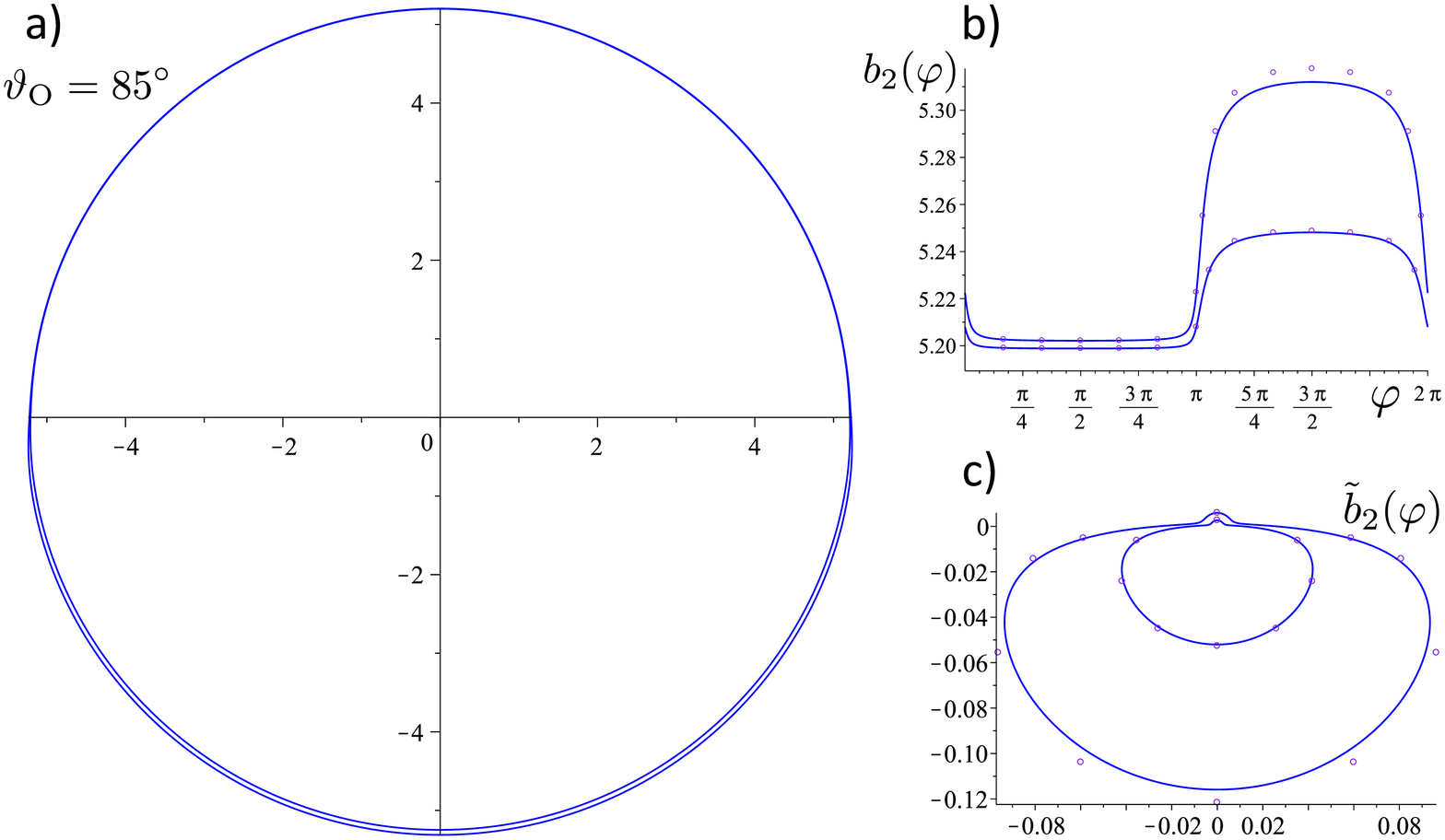}
\end{center}
\caption{Thickness and shape of tertiary ($n=2$) ring image of luminous accretion disk with inner and outer boundaries $r_S^\mathrm{in}= 6m$ and $r_S^\mathrm{out}= 30m$, as seen by nearly equatorial observer, $\vartheta_\mathrm{O}=85^\circ$. All graphs are based on the analytical formula (\ref{b-varphi}). Panel 'a' shows the photon ring in real scale. The shape of the curve is almost indistinguishable from a circle (we remind that the shadow of Schwarzschild black hole is a circle with the radius $b_{cr}=3\sqrt{3}m \simeq 5.196m$). However, it is still possible to notice the thickness of the ring at the lower half of the image. Panel 'b' shows the impact parameter $b_2$ in the observer's sky as the function of $\varphi$. Panel 'c' shows the polar plot of the excess $\tilde{b}_2(\varphi)$, defined by eq.(\ref{excess}). The small circles show the results of numerical calculations, see Subsection \ref{subsec-errors} for details.}
\label{fig:theta-85}
\end{figure*}

\section{Discussion}
\label{sec:discussion}

In this Section, we present graphs for the higher-order photon rings based on our formula, and derive some analytical properties of the rings.

\subsection{Shape of photon rings}
\label{subsec-shape}

First of all, our analytical formula allows us to plot graphs showing the shape of photon rings. The Fig.\ref{fig:th-several} shows graphs for different radii of the emitting circle and different angles of inclination of the observer. It should be kept in mind that the difference between the shape of the higher-order rings and the shadow boundary is difficult to notice by looking at the ring itself, because the higher-order images are exponentially close to the shadow edge. In particular, for the Schwarzschild black hole considered here, the shape of the higher-order rings differs very little from the circle.

To demonstrate the shape of the $n=2$ ring, in Fig.\ref{fig:th-several} we plot the impact parameter $b_2$ as the function of $\varphi$, where $\varphi$ is presented on the horizontal axis (left panels).

In addition, for a better understanding of the shape, we use the following technique: instead of the plotting the impact parameter itself, we make the polar plot of the 'excess'
\begin{equation} \label{excess}
\tilde{b}_2(\varphi) = b_2(\varphi) - 3\sqrt{3}m   \, .   
\end{equation}
Such a graph shows the radial offset of the image point from the shadow boundary in each polar direction, see right panels in Fig.\ref{fig:th-several}.

In the graph for the angle $\vartheta_\mathrm{O}=17^\circ$ in Fig.\ref{fig:th-several}, one can notice a sinusoidal oscillation of the impact parameter; see also Fig.7 in \cite{Gralla-Lupsasca-Marrone-2020} for Kerr case. Having an analytical formula for the curve allows us to derive this property analytically. Expanding eq.(\ref{b-varphi-general-n}) in small angle $\vartheta_\mathrm{O}$, we find:
\[
b_n(\varphi) \simeq 3\sqrt{3}m \, \big\{ 1 +  f(r_S) \, e^{-(n+1)\pi/2} \, \times
\big. 
\]
\begin{equation} \label{b-varphi-polar}
\big. \times \, (1 - \vartheta_\mathrm{O} \sin \varphi ) \big\} \, .
\end{equation}
As can be seen, the linear term of the expansion is proportional to $\sin \varphi$. We conclude that for nearly polar observer, $\vartheta_\mathrm{O} \ll 1$, the ring size oscillates sinusoidally.

We also mention that for polar observer (zero inclination, $\vartheta_\mathrm{O} = 0$), the higher-order images have the circular shape with radius (in terms of impact parameter):
\begin{equation} \label{b-previous}
b_n = 3\sqrt{3}m \left\{ 1 +  f(r_S) \, e^{-(n+1)\pi/2} \right\} \, ,
\end{equation}
which agrees with the results of our previous paper \cite{BK-Tsupko-2022}.


\subsection{Thickness of photon rings}
\label{subsec-thickness}

In order to discuss the thickness (\ref{thickness-general}) of the photon rings, consider a thin luminous accretion disk with given inner and outer radii, $r_S^\mathrm{in}$ and $r_S^\mathrm{out}$.

Let us discuss the tertiary image of such a disk, as viewed at different inclinations.
If the observer looks at the disk with zero inclination (polar view, $\vartheta_\mathrm{O} = 0$), all parts of the ring will have the same thickness. All of them will be formed by photons for which the change of angular coordinate during the motion to observer is equal to $\Delta \tilde{\phi} = 5\pi/2$, see (\ref{b-previous}) with $n=2$. In this case, the thickness is so small that it is rather difficult to show in the picture, see Fig.7 in \cite{BK-Tsupko-2022}.

To make the thickness of the ring on the graph more noticeable, it is better to consider the observer at a high inclination (close to the equatorial view, $\vartheta_\mathrm{O} \lesssim \pi/2$). For such observer, different parts of the photon ring will have quite different thicknesses. The lower parts of the image will have the biggest thickness because they are formed by photons with a smaller total bending. Namely, for the tertiary ring, the lowest points of the image are formed by photons with $\Delta \tilde{\phi} \sim 2\pi$ and the upper points of the image will have $\Delta \tilde{\phi} \sim 3\pi$.

Based on the discussion above, in Fig.\ref{fig:theta-85}a we try to show the thickness of the ring on a real scale, for the case when observer is close to equatorial plane and, correspondingly, the lower part of the ring is relatively thick. The dependence of thickness on the polar direction can be understood from Fig.\ref{fig:theta-85}b,c. \\

\begin{figure}
\begin{center}
\includegraphics[width=0.40\textwidth]{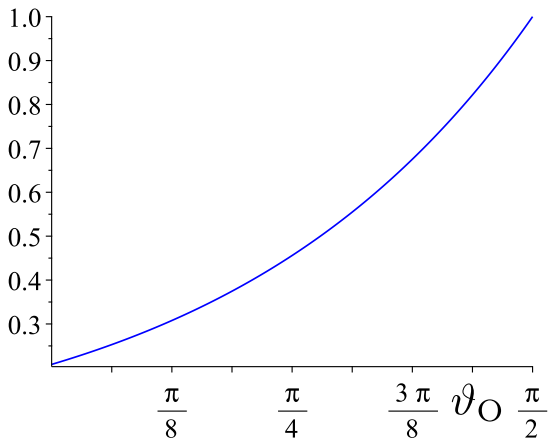}
\end{center}
\caption{Right-hand side of Eq.\ref{eq-thickness} as the function of $\vartheta_\mathrm{O}$.}
\label{fig:thickness}
\end{figure}

The higher-order ring image of accretion disk of given size has the maximum thickness at $\varphi=3\pi/2$, see Eq.(\ref{thickness-general}). The maximum thickness of the $n$-ring $(\Delta b_n)_\mathrm{max}$ depends on the inclination $\vartheta_\mathrm{O}$ as
\begin{equation} 
(\Delta b_n)_\mathrm{max}  \propto  \exp \left[  \arccos \left( \frac{-1}{\sqrt{1 + \cot^2 \vartheta_\mathrm{O}} } \right) \right] \, ,
\end{equation}
or, using $\arccos(-x)=\pi - \arccos{x}$, as
\begin{equation} \label{eq-thickness}
(\Delta b_n)_\mathrm{max}  \propto  \exp \left[ - \arccos \left( \frac{1}{\sqrt{1 + \cot^2 \vartheta_\mathrm{O}} } \right) \right] \, .
\end{equation}
In Fig.\ref{fig:thickness}, we present the graph of the right-hand side of the Eq.(\ref{eq-thickness}). We can also conclude that the observation of the tertiary ring might be most promising for the observer located slightly above the equatorial plane.

A discussion of the accuracy of the ring thickness calculation is given in the Subsection \ref{subsec-errors}.

\subsection{Area occupied by the ring}
\label{subsec-area}

Let us calculate the area (\ref{area-general}) of $n$-th photon ring, as viewed at the small inclination, $\vartheta_\mathrm{O} \ll 1$. Up to the linear order in angle $\vartheta_\mathrm{O}$, we find:
\begin{equation} \label{area-small-inclination}
\Delta S_n \simeq 54 \pi m^2  [f(r_S^\mathrm{out}) - f(r_S^\mathrm{in})] e^{-(n+1)\pi/2} \, .
\end{equation}
The term, linear in $\vartheta_\mathrm{O}$ in (\ref{b-varphi-polar}), gives zero contribution when integrated over $\varphi$.

We can conclude that, for nearly polar observer, the area occupied by the image does not depend on the inclination of observer (up to the first order in $\vartheta_\mathrm{O}$). This means that some of the results obtained in the paper \cite{BK-Tsupko-2022} for zero inclination can be used to make estimations for the case of nonzero inclination as well.

A numerical example of the accuracy of the area calculation is given in the Subsection \ref{subsec-errors}.

\begin{figure}
\begin{center}
\includegraphics[width=0.45\textwidth]{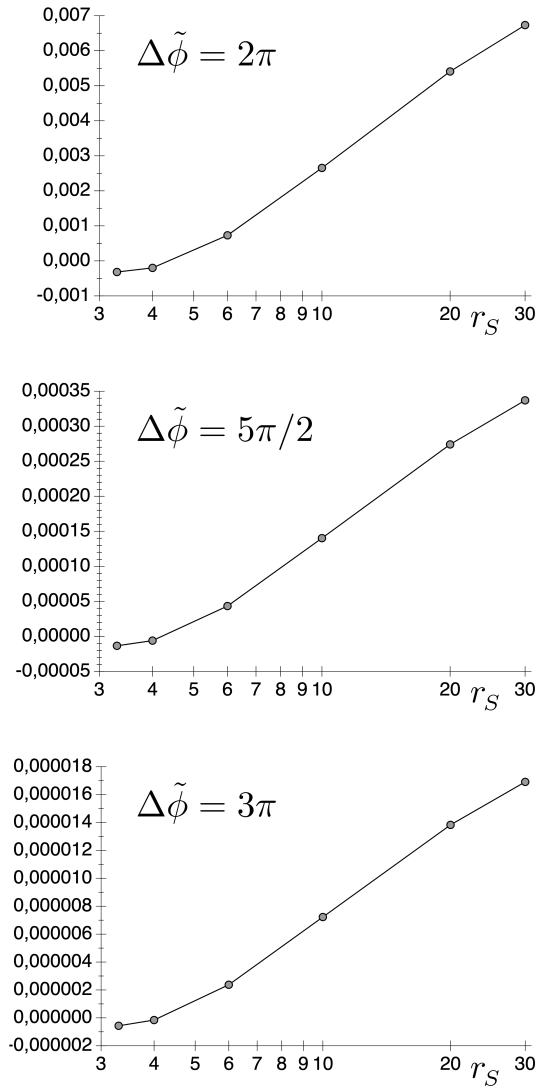}
\end{center}
\caption{Difference between the exact value of $b$ and analytical approximation of $b$ by Eq.(\ref{b-general}), in units of $m$ (absolute error), as function of $r_S$, plotted for three values of photon angular shift $\Delta \tilde{\phi} = 2 \pi, \, 5\pi/2, \, 3 \pi$. Since the analytical calculations used here differ only in the value of the angle $\Delta \tilde{\phi}$ in the exponent in Eq.(\ref{b-general}), then the shape of curves is approximately the same. What matters here is the different scale on the vertical axes. Scale of errors can be compared with scales of Fig.\ref{fig:th-several},\ref{fig:theta-85}. See also Fig.4b in \cite{Bozza-Scarpetta-2007}.}
\label{fig:errors}
\end{figure}

\subsection{Accuracy of approximations}
\label{subsec-errors}

The accuracy of our approximate formula (\ref{b-varphi-general-n}) and other formulas and the graphs obtained from it are entirely determined by the accuracy of the strong deflection limit approximation (\ref{bozza-2007-03}),(\ref{bozza-2007-04}),(\ref{f-rS}),(\ref{b-general}). All other geometric relations used are exact.

In this paper, we consider only higher-order images ($n \ge 2$). For such images, the strong deflection approximation is known to have a high accuracy \cite{Bozza-Scarpetta-2007, Bozza-2010, Aldi-Bozza-2017}. Below we will discuss the tertiary $n=2$ image only. For images of higher orders than $n=2$, the strong deflection approximation works even better.

Calculating the exact values of the impact parameter is straightforward. For a given value $r_S$ of the initial point, the change in the angular coordinate $\Delta \tilde{\phi}$ of the ray for a known impact parameter $b$ is found by integrating the orbit equation of light ray within the required limits. In our case, we first choose the emission ring radius $r_S$ and the angle of the observer's inclination $\vartheta_\mathrm{O}$. Then, for every value of angle $\varphi$, one can find $\Delta \tilde{\phi}$ by formulas (\ref{shift-01}),(\ref{gamma-alpha}),(\ref{alpha-to-varphi}). Knowing the value of $\Delta \tilde{\phi}$, one can numerically find the corresponding impact parameter $b$.

In Figs.\ref{fig:th-several},\ref{fig:theta-85}, together with analytical curves, we plot the numerical points of $b$. Since each pair of left and right panels are drawn for the same case, and also have about the same scale, one can get a good idea of the calculation errors. For example, to understand the behavior of the upper parts of the curve for $\vartheta_\mathrm{O}=85^\circ,60^\circ$, it is better to look at the left-hand graphs, because on the right-hand graphs this part is very compacted. In Fig.\ref{fig:th-several}, the difference is noticeable only for large values of the radius $r_S$ and for a large inclination $\vartheta_\mathrm{O}$ of the observer. For an observer close to polar (lower pair of panels in Fig.\ref{fig:th-several}), the numerical points are almost indistinguishable from the analytical ones.\\

Additionally, to demonstrate the magnitude of the errors in the Figs.\ref{fig:th-several},\ref{fig:theta-85} more clearly, we will present the following illustrative example. As already discussed in Subsection \ref{subsec-thickness}, for a polar observer, all photons forming the tertiary image experience the same change of angular coordinate equal to $\Delta \tilde{\phi} = 5\pi/2$. At the same time, for nearly equatorial observer ($\vartheta_\mathrm{O} \lesssim \pi/2$), different parts of the tertiary image will be formed by photons with very different values of $\Delta \tilde{\phi}$, from $\Delta \tilde{\phi} \sim 2\pi$ to $\Delta \tilde{\phi} \sim 3\pi$. Therefore, as numerical example, we find and plot the error of approximation of $b$ for three values of the azimuthal shift: $\Delta \tilde{\phi} = 2 \pi, \, 5\pi/2, \, 3 \pi$. We take the values of the radius $r_S$ of the emitting circular ring the same as used in the Figs.\ref{fig:th-several},\ref{fig:theta-85}, see Fig.\ref{fig:errors}.

According to the discussion above, panels in Fig.\ref{fig:errors} cover the whole range of values that the error of calculating the impact parameter for the tertiary image may have, within all possible angles of inclination: the upper graph can be considered as the maximum possible error, the middle one shows the typical error, and the lower one can be interpreted as the minimum error.

Also, it is clear that for the $n=3$ and all subsequent images the error in all pictures will be absolutely negligible, because for such curves the change of angle for all points will be $\Delta \tilde{\phi} > 3 \pi$. \\

\begin{figure}
\begin{center}
\includegraphics[width=0.48\textwidth]{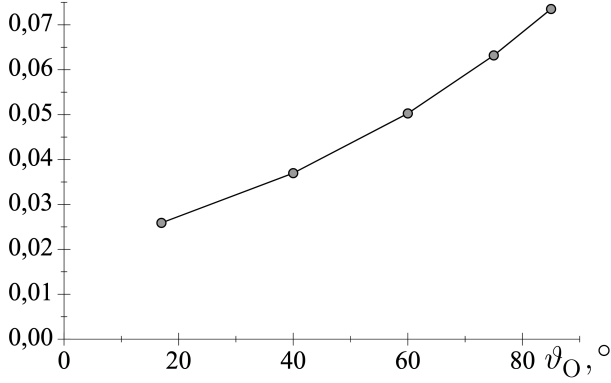}
\end{center}
\caption{Relative error of calculation of maximum thickness of the $n=2$ image of the accretion disk with inner and outer radii, $r_S^\mathrm{in}=6m$ and $r_S^\mathrm{out}=30m$, as the function of the inclination angle $\vartheta_\mathrm{O}$. The following values are used: $\vartheta_\mathrm{O} = 17^\circ, 40^\circ, 60^\circ, 75^\circ, 85^\circ$ (compare with Fig.\ref{fig:th-several}).}
\label{fig:error-thickness}
\end{figure}

In order to illustrate the accuracy of the ring thickness calculations (Subsection \ref{subsec-thickness}) and, in particular, of Eq.(\ref{eq-thickness}), we present the following numerical example. We consider the accretion disk with the given inner and outer radii the same as in Fig.\ref{fig:theta-85}. For this disk, we calculate the width of the thickest part $(\Delta b_2)_\mathrm{max}$ of the tertiary ring, numerically and analytically. In the Fig.\ref{fig:error-thickness} we show the relative error of calculating the maximum thickness of the given ring, as the function of observer's inclination $\vartheta_\mathrm{O}$. As one can see, the error decreases as the observer's inclination decreases.

For estimation of the accuracy of area calculation (Subsection \ref{subsec-area}), we consider the same disk ($r_S^\mathrm{in}=6m$ and $r_S^\mathrm{out}=30m$), viewed face-on ($\vartheta_\mathrm{O}=0$). In this simplest case, the area $\Delta S_2$ of its tertiary image can be calculated as $\pi (b_\mathrm{out}^2 - b_\mathrm{in}^2 )$ numerically and analytically. We find that the accuracy is about 2.0$\%$.

\section{Conclusions}
\label{sec:conslusion} 

(i) Studies of higher-order photon rings, especially tertiary (number of half-orbits $n=2$) and quaternary ($n=3$), are of great interest in the perspective of future observations. In this regard, it would be very useful to have an analytical description of the shape of these rings, even for simple case.

(ii) In this paper, we derive the equation describing the shape of higher-order ring as the polar curve: the impact parameter as the function of the angle on the observer's sky. Our formula describes the apparent shape of lensed image of circular ring with given radius around the Schwarzschild black hole, for an arbitrary observer's inclination (Fig.\ref{fig:geometry}). The analytical expression for any higher-order ring $n \ge 2$ is given by Eq.(\ref{b-varphi-general-n}). The summary of all variables used is given after Eq.(\ref{b-varphi-general-n}).

(iii) Our formula is a simple and efficient alternative to the numerical calculation of ray trajectories. First, it allows one to easily plot the higher-order ring curves. As an example of the use of our formula, we have presented graphs illustrating the shape of the tertiary ring (Figs.\ref{fig:th-several},\ref{fig:theta-85}). Second, the analytical expression allows one to derive a number of universal properties without the need for numerical simulations. As an example, we derived analytically some properties of rings, see Section \ref{sec:discussion}.

\section*{Acknowledgements}

This work is supported by the Russian Science Foundation, Grant No. 18-12-00378.
Author is thankful to V. Perlick for useful discussions.
Author brings special thanks to G.S. Bisnovatyi-Kogan for motivation and permanent help with all scientific initiatives.

\section*{Appendix A. Derivation from results of Aldi and Bozza}

In this Appendix we show how to obtain our formula from the results of Aldi and Bozza \cite{Aldi-Bozza-2017}. In their work, the authors have found an analytical relationship between the higher-order image parameters in the observer's sky and the position of the emitting source in the equatorial plane. For every image order $n$, the two halves of the accretion disk are considered separately, with the corresponding choice of the parameter $m$ (should not be confused with our mass parameter) and the additional parameter $\sigma$ (see below). We have succeeded to show that for our problem these two 'branches' of solution can be combined into one formula containing only the image order $n$. For simplicity, we compare only $n=2$ and $n=3$ rings. The full agreement is found.

According to \cite{Aldi-Bozza-2017}, the variables ($\epsilon, \xi$) specifying the position of the image element on the observer's sky are associated with variables ($r_e, \phi_e$) specifying the position of the source element as
\begin{equation} \label{xi-Bozza}
\xi = - \sigma \frac{\tan \phi_e}{\sqrt{\mu_\mathrm{O}^2 + \tan^2 \phi_e}} \, ,
\end{equation}
and
\begin{equation} \label{epsilon-Bozza}
\epsilon = \bar{\epsilon}(r_e) \, \hat{\epsilon}(\phi_e) \, ,
\end{equation}
where the function $\bar{\epsilon}(r_e)$ agrees with our function $f(r_S)$ (we will show it later), and the function $\hat{\epsilon}(\phi_e)$ is given by
\begin{equation} \label{hat-epsilon-Bozza}
\hat{\epsilon}(\phi_e) = \exp \left[-m \pi - \sigma \arcsin \sqrt{\mu_\mathrm{O}^2 \cos^2 \phi_e + \sin^2 \phi_e  } \right] \, .
\end{equation}
Here $m$ is the number of polar inversions of the photon, which depends on the image order $n$ and the considered half of the disk, see the explanations after Eq.(3.28) in \cite{Aldi-Bozza-2017}. In turn, the parameter $\sigma$ is
\begin{equation} \label{sigma-definition}
\sigma = \pm (-1)^m \, .
\end{equation}
The choice of sign in $\sigma$ is explained after Eq.(3.21) in \cite{Aldi-Bozza-2017}. One needs to choose a positive sign for photons emitted upwards from the disk and negative for photons emitted downwards \cite{Aldi-Bozza-2017}. Variable $\mu_\mathrm{O}$ defines the observer inclination: $\mu_\mathrm{O} \equiv \cos \vartheta_\mathrm{O}$. Also, the position angle in the observer sky $\varphi$ (note that authors \cite{Aldi-Bozza-2017} use another letter here) is given by
\begin{equation} \label{varphi-Bozza}
\varphi = - \sigma \arccos(-\xi) \, .
\end{equation}

We start from $n=2$ photon ring. First, we rewrite our eq.(\ref{b-varphi}) as
\begin{equation} \label{epsilon-mine}
\epsilon = f(r_S) \,  \exp \left[ -3\pi + \arccos \left( \frac{\sin \varphi}{\sqrt{\sin^2 \varphi + \cot^2 \vartheta_\mathrm{O}} } \right) \right]   \, .
\end{equation}

For $n=2$ ring, with our geometry, we need to choose 'plus' sign in $\sigma$ for this image.

Authors \cite{Aldi-Bozza-2017} put the observer to the position $\phi_\mathrm{O}=\pi$ and consider two halves of the accretion disk separately. First, consider the half of the disk closer to the observer, $\pi/2 < \phi_e < 3\pi/2$. According to discussion after eq.(3.28) in \cite{Aldi-Bozza-2017}, for image of $n=2$ order, one needs to take $m=2$. Therefore, we find $\sigma=+1$.

With these values of $m$ and $\sigma$, Eqs. (\ref{varphi-Bozza}) and (\ref{xi-Bozza}) lead to:
\begin{equation}
\varphi = - \arccos  \frac{\tan \phi_e}{\sqrt{\mu_\mathrm{O} ^2 + \tan^2 \phi_e}}  \, .
\end{equation}
Using the relation $\sin(\arccos x) = \sqrt{1-x^2}$, one can get:
\begin{equation} \label{neg-sin-Bozza}
\sin \varphi = - \frac{\mu_\mathrm{O}}{\sqrt{\mu_\mathrm{O} ^2 + \tan^2 \phi_e}}  < 0  \, .
\end{equation}

Now we transform $\arcsin$ to $\arccos$ in the formula (\ref{hat-epsilon-Bozza}), using the relation $\arcsin x = \arccos \sqrt{1-x^2}$:
\[
\arcsin \sqrt{\mu_\mathrm{O}^2 \cos^2 \phi_e + \sin^2 \phi_e  } =
\]
\begin{equation} \label{arcsin-arccos-Bozza}
= \arccos \sqrt{ (1 - \mu_\mathrm{O}^2 ) \cos^2 \phi_e  }  \, .
\end{equation}
We need to express $\cos^2 \phi_e$ via $\varphi$. Squaring the equation (\ref{neg-sin-Bozza}), we get
\begin{equation}
\sin^2 \varphi = \frac{\mu_\mathrm{O}^2}{\mu_\mathrm{O} ^2 + \tan^2 \phi_e}   \, .
\end{equation}
Rearranging the terms, we find:
\begin{equation} \label{tan11-Bozza}
\tan^2 \phi_e = \mu_\mathrm{O}^2 \cot^2 \varphi \, .
\end{equation}
Expressing $\cos^2 \phi_e$ via $\tan^2 \phi_e$, with help of (\ref{tan11-Bozza}), we find:
\begin{equation} \label{cos11-Bozza}
\cos^2 \phi_e = \frac{1}{1 + \mu_\mathrm{O}^2 \cot^2 \varphi} \, .
\end{equation}
Substituting expression (\ref{cos11-Bozza}) into Eq.(\ref{arcsin-arccos-Bozza}), after some transformations, we find:
\[
\arcsin \sqrt{\mu_\mathrm{O}^2 \cos^2 \phi_e + \sin^2 \phi_e  } =
\]
\begin{equation} \label{Bozza-123}
= \arccos \sqrt{\frac{\sin^2 \varphi}{\sin^2 \varphi + \cot^2 \vartheta_\mathrm{O}}     }  \, .
\end{equation}

According to Eq.(\ref{neg-sin-Bozza}), for $\pi/2 < \phi_e < 3\pi/2$, we have $\sin \varphi < 0$. Therefore,
\begin{equation}
\sqrt{\sin^2 \varphi} = - \sin \varphi \, ,
\end{equation}
and finally we find:
\[
\arcsin \sqrt{\mu_\mathrm{O}^2 \cos^2 \phi_e + \sin^2 \phi_e  } =
\]
\begin{equation} \label{arcsin-arccos-Bozza-2}
= \arccos \frac{- \sin \varphi }{ \sqrt{ \sin^2 \varphi + \cot^2 \vartheta_\mathrm{O}}    } \, .
\end{equation}

Using (\ref{arcsin-arccos-Bozza-2}) in (\ref{hat-epsilon-Bozza}), together with $m=2$ and $\sigma=+1$, we find:
\begin{equation}
\hat{\epsilon}(\phi_e) = \exp \left[ -2\pi - \arccos \frac{ - \sin \varphi }{ \sqrt{ \sin^2 \varphi + \cot^2 \vartheta_\mathrm{O}}    }  \right]  \, .
\end{equation}
Using the relation $\arccos(-x)=\pi - \arccos{x}$, we find the final expression for the function $\hat{\epsilon}(\phi_e)$:
\begin{equation} \label{hat-epsilon-final}
\hat{\epsilon}(\phi_e) = \exp \left[ -3\pi + \arccos \frac{ \sin \varphi }{ \sqrt{ \sin^2 \varphi + \cot^2 \vartheta_\mathrm{O}}    }  \right]  \, ,
\end{equation}
which agrees with our Eq.(\ref{epsilon-mine}).

Now we will consider the photons emitted from the far side of the disk, where $-\pi/2 < \phi_e < \pi/2$. According to explanations after eq.(3.28) in \cite{Aldi-Bozza-2017}, for image of $n=2$ order, one needs to take $m=3$. Therefore, we find $\sigma=-1$. With these values of $m$ and $\sigma$, Eqs. (\ref{varphi-Bozza}) and (\ref{xi-Bozza}) lead to:
\begin{equation}
\varphi = + \arccos  \frac{- \tan \phi_e}{\sqrt{\mu_\mathrm{O} ^2 + \tan^2 \phi_e}}  \, .
\end{equation}
From this, one can get
\begin{equation} \label{pos-sin-Bozza}
\sin \varphi = + \frac{\mu_\mathrm{O}}{\sqrt{\mu_\mathrm{O} ^2 + \tan^2 \phi_e}}  > 0   \, .
\end{equation}

Part of transformation of $\arcsin$ with the square root in Eq.(\ref{hat-epsilon-Bozza}) will be exactly the same, as for previous range of $\phi_e$, and we reach the formula (\ref{Bozza-123}). But now, according to (\ref{pos-sin-Bozza}), we have $\sin \varphi >0$. Keeping in mind that the values of $m$ and $\sigma$ have also changed, we finally come to the same Eq.(\ref{hat-epsilon-final}).

Thus, we have shown that with the help of our transformations, formulas for two ranges of angle $\phi_e$ can be combined into one expression (\ref{hat-epsilon-final}), which is in agreement with our Eq.(\ref{epsilon-mine}).\\

Now we consider $n=3$ photon ring. With our geometry, we need to choose ’minus’ sign in (\ref{sigma-definition}) for this image.

First, consider the half of the disk closer to the observer, $\pi/2 < \phi_e < 3\pi/2$. According to discussion after eq.(3.28) in \cite{Aldi-Bozza-2017}, for image of $n=3$ order, one needs to take $m=4$. Therefore, we find $\sigma=-1$. We obtain:
\begin{equation}
\varphi = + \arccos  \frac{-\tan \phi_e}{\sqrt{\mu_\mathrm{O} ^2 + \tan^2 \phi_e}}  \, , \quad \sin \varphi > 0 \, .
\end{equation}
Analogous to the transformations done for the $n=2$ ring, we find:
\begin{equation} \label{hat-epsilon-final-n3}
\hat{\epsilon}(\phi_e) = \exp \left[ -4\pi + \arccos \frac{ \sin \varphi }{ \sqrt{ \sin^2 \varphi + \cot^2 \vartheta_\mathrm{O}}    }  \right]  \, .
\end{equation}

For the photons emitted from the far side of the disk, where $-\pi/2 < \phi_e < \pi/2$, we have $m=3$ and $\sigma=+1$. Correspondingly, we obtain:
\begin{equation}
\varphi = - \arccos  \frac{\tan \phi_e}{\sqrt{\mu_\mathrm{O} ^2 + \tan^2 \phi_e}}  \, , \quad \sin \varphi < 0 \, .
\end{equation}
Analogous to the transformations done for the $n=2$ ring, we find the same Eq.(\ref{hat-epsilon-final-n3}) for these range of $\phi_e$ too. Eq.(\ref{hat-epsilon-final-n3}) agrees with our Eq.(\ref{b-varphi-n3}).\\

Now we will show that our function $f(r_S)$ is equivalent to $\bar{\epsilon}(r_e)$ used in eq.(\ref{epsilon-Bozza}) and defined in eq.(3.23) of \cite{Aldi-Bozza-2017}. 
We rewrite eq.(\ref{f-rS}) as:
\begin{equation}
f(r_S) = \frac{6^5 (r_S - 3m)}{ (3+\sqrt{3})^2 (3\sqrt{r_S} + \sqrt{3r_S + 18m})^2 } =
\end{equation}
\[
= \frac{216 \cdot 4 \, (r_S - 3m)}{ (1+\sqrt{3})^2 (\sqrt{3r_S} + \sqrt{r_S + 6m})^2 } \, .
\]
Using the following relations,
\begin{equation}
(1+\sqrt{3})^2 = 2 \, (2+\sqrt{3}) \, ,
\end{equation}
\begin{equation}
\frac{1}{2 + \sqrt{3}} = 2 - \sqrt{3} \, ,
\end{equation}
\begin{equation}
(\sqrt{3r_S} + \sqrt{r_S + 6m})  (\sqrt{3r_S} - \sqrt{r_S + 6m})  =  
\end{equation}
\[
= 2 \, (r_S - 3m) \, ,
\]
we find
\begin{equation} \label{Bozza-125}
f(r_S) = \frac{216 \, (2-\sqrt{3})  (\sqrt{3r_S} - \sqrt{r_S + 6m})}{ \sqrt{3r_S} + \sqrt{r_S + 6m} } \, .
\end{equation}
This expression agrees with eq.(3.23) of \cite{Aldi-Bozza-2017} if one change $r_S \to r_e$. Note that authors \cite{Aldi-Bozza-2017} use the units of the Schwarzschild radius, $2MG/c^2 = 1$.

\bibliographystyle{ieeetr}

\begin{thebibliography}{11}
	


\bibitem{Perlick-2004a}
V. Perlick,
Gravitational lensing from a space-time Perspective,
Liv. Rev. Relativity, 7, 9 (2004)




\bibitem{Bozza-2010}
V. Bozza,
Gravitational lensing by black holes,
Gen. Rel. Grav. 42, 2269 (2010)




\bibitem{BK-Tsupko-Universe-2017}
G. S. Bisnovatyi-Kogan and O. Yu. Tsupko,
Gravitational lensing in presence of plasma: Strong lens systems, black hole lensing and shadow,
Universe 3, 57 (2017) 





\bibitem{BK-Tsupko-2022}
G.S. Bisnovatyi-Kogan and O.Yu. Tsupko,
Analytical study of higher-order ring images of the accretion disk around a black hole,
Phys. Rev. D 105, 064040 (2022)

	

\bibitem{Darwin-1959}
Ch. Darwin,
The gravity field of a particle,
Proceedings of the Royal Society of London. Series A, Mathematical and Physical Sciences, 249, 180 (1959)


\bibitem{Atkinson-1965}
R. Atkinson,
On light tracks near a very massive star,
Astronomical Journal, 70, 517 (1965)



\bibitem{MTW-1973}
C.W. Misner, K.S. Thorne, J.A. Wheeler, \textit{Gravitation},
W.H. Freeman and Company, San Francisco (1973)



\bibitem{Ohanian-1987}
H. C. Ohanian,
The black hole as a gravitational ``lens'',
American Journal of Physics 55, 428 (1987)


\bibitem{Bao-Hadrava-1994a}
G. Bao, P. Hadrava, E. Ostgaard,
Multiple Images and Light Curves of an Emitting Source on a Relativistic Eccentric Orbit around a Black Hole,
Astrophysical Journal, 425, 63 (1994)


\bibitem{Bao-Hadrava-1994b}
G. Bao, P. Hadrava, E. Ostgaard,
Emission-Line Profiles from a Relativistic Accretion Disk and the Role of Its Multiple Images,
Astrophysical Journal, 435, 55 (1994)





\bibitem{Virbhadra-2000}
K.S. Virbhadra, G.F.R. Ellis,
Schwarzschild black hole lensing,
Phys. Rev. D 62 (2000) 084003




\bibitem{Virbhadra-2001}
K. S. Virbhadra and G. F. R. Ellis,
Gravitational lensing by naked singularities,
Phys. Rev. D 65, 103004 (2002) 



\bibitem{Virbhadra-2009}
K. S. Virbhadra,
Relativistic images of Schwarzschild black hole lensing,
Phys. Rev. D 79, 083004 (2009)


\bibitem{Bozza-2001}
V. Bozza, S. Capozziello, G. Iovane, G. Scarpetta,
Strong field limit of black hole gravitational lensing,
Gen. Rel. Grav. 33, 1535 (2001) 




\bibitem{Bozza-2002}
V. Bozza,
Gravitational lensing in the strong field limit,
Phys. Rev. D 66, 103001 (2002) 




\bibitem{Bozza-2003}
V. Bozza,
Quasiequatorial gravitational lensing by spinning black holes in the strong field limit,
Phys. Rev. D 67, 103006 (2003)





\bibitem{Bozza-Mancini-ApJ-2004}
V. Bozza and L. Mancini,
Gravitational Lensing by Black Holes: A Comprehensive Treatment and the Case of the Star S2,
The Astrophysical Journal, 611, 1045 (2004)


\bibitem{Bozza-Mancini-ApJ-2005}
V. Bozza and L. Mancini,
Gravitational Lensing of Stars in the Central Arcsecond of Our Galaxy,
The Astrophysical Journal, 627, 790 (2005)




\bibitem{Bozza-2005}
V. Bozza, F. De Luca, G. Scarpetta, M. Sereno,
Analytic Kerr black hole lensing for equatorial observers in the strong deflection limit,
Phys. Rev. D, 72, 083003 (2005)


\bibitem{Bozza-2006}	
V. Bozza, F. De Luca, G. Scarpetta,
Kerr black hole lensing for generic observers in the strong deflection limit,
Phys. Rev. D 74, 063001 (2006) 



\bibitem{Bozza-Sereno-2006}
V. Bozza and M. Sereno,
Weakly perturbed Schwarzschild lens in the strong deflection limit,
Phys. Rev. D 73, 103004 (2006)




\bibitem{Bozza-2008}
V. Bozza,
Comparison of approximate gravitational lens equations and a proposal for an improved new one,
Phys. Rev. D 78, 103005 (2008)




\bibitem{Eiroa-2002}
E.F. Eiroa, G.E. Romero, and D.F. Torres,
Reissner-Nordstr\"{o}m black hole lensing,
Phys. Rev. D 66, 024010 (2002)



\bibitem{Eiroa-2004}
E.F. Eiroa and D.F. Torres,
Strong field limit analysis of gravitational retrolensing,
Phys. Rev. D 69, 063004 (2004)



\bibitem{Eiroa-2005}
E.F. Eiroa,
Braneworld black hole gravitational lens: Strong field limit analysis,
Phys. Rev. D 71, 083010 (2005)



\bibitem{Eiroa-2006}
E.F. Eiroa,
Gravitational lensing by Einstein-Born-Infeld black holes,
Phys. Rev. D 73, 043002 (2006)




\bibitem{Frittelli-2000}
S. Frittelli, Th.P. Kling, and E.T. Newman,
Spacetime perspective of Schwarzschild lensing,
Phys. Rev. D 61, 064021 (2000)





\bibitem{Perlick-2004b}
V. Perlick,
Exact gravitational lens equation in spherically symmetric and static space-times,
Phys. Rev. D, 69, 064017 (2004) 








\bibitem{Amore-2007}
P. Amore, M. Cervantes, A. De Pace, and Francisco M. Fern\'{a}ndez,
Gravitational lensing from compact bodies: Analytical results for strong and weak deflection limits,
Phys. Rev. D 75, 083005 (2007)



\bibitem{Iyer-Petters-2007}
S. V. Iyer and A. O. Petters,
Light’s bending angle due to black holes: from the photon sphere to infinity,
General Relativity and Gravitation, 39, 1563 (2007)



\bibitem{Gyulchev-2007}
G.N. Gyulchev and St.S. Yazadjiev,
Kerr-Sen dilaton-axion black hole lensing in the strong deflection limit,
Phys. Rev. D 75, 023006 (2007)



\bibitem{BK-Tsupko-2008}
G.S. Bisnovatyi-Kogan, O.Yu. Tsupko,
Strong gravitational lensing by Schwarzschild black holes,
Astrophysics 51, 99 (2008)



\bibitem{Chen-Jing-2009}
S. Chen and J. Jing,
Strong field gravitational lensing in the deformed Ho\v{r}ava-Lifshitz black hole,
Phys. Rev. D 80, 024036 (2009)






\bibitem{Ghosh-2010}
T. Ghosh and S. SenGupta,
Strong gravitational lensing across a dilaton anti–de Sitter black hole,
Phys. Rev. D 81, 044013 (2010)







\bibitem{Bin-Nun-2011}
A. Y. Bin-Nun,
Strong gravitational lensing by Sgr A*,
Classical and Quantum Gravity, 28, 114003 (2011).




\bibitem{Ding-Kang-2011}
Ch. Ding, Sh. Kang, Ch.-Y. Chen, S. Chen, and J. Jing,
Strong gravitational lensing in a noncommutative black-hole spacetime,
Phys. Rev. D 83, 084005 (2011)






\bibitem{Wei-2012}
Sh.-W. Wei, Y.-X. Liu, Ch.-E Fu and K. Yang,
Strong field limit analysis of gravitational lensing in Kerr-Taub-NUT spacetime,
Journal of Cosmology and Astroparticle Physics, 10, 053 (2012).



\bibitem{Tsupko-BK-2013}
O. Yu. Tsupko, G. S. Bisnovatyi-Kogan,
Gravitational lensing in plasma: Relativistic images at homogeneous plasma,
Phys. Rev. D 87, 124009 (2013)



\bibitem{Sadeghi-2014}
J. Sadeghi and H. Vaez,
Strong gravitational lensing with Gauss-Bonnet correction,
Journal of Cosmology and Astroparticle Physics, 06, 028, (2014)


\bibitem{Alhamzawi-2016}
A. Alhamzawi and R. Alhamzawi,
Gravitational lensing in the strong field limit by modified gravity,
General Relativity and Gravitation, 48, 167 (2016)


\bibitem{Tsukamoto-2016}
N. Tsukamoto,
Strong deflection limit analysis and gravitational lensing of an Ellis wormhole,
Phys. Rev. D 94, 124001 (2016)



\bibitem{Chakraborty-2017}
S. Chakraborty and S. SenGupta,
Strong gravitational lensing—a probe for extra dimensions and Kalb-Ramond field,
Journal of Cosmology and Astroparticle Physics, 07, 045 (2017)




\bibitem{Barlow-2017}
N.S. Barlow, S.J. Weinstein and J.A. Faber,
An asymptotically consistent approximant for the equatorial bending angle of light due to Kerr black holes,
Classical and Quantum Gravity, 34, 135017 (2017).





\bibitem{Bozza-2017}
V. Bozza,
Gravitational lensing by black holes and their alternatives,
International Journal of Modern Physics D, 26, 1741013 (2017)





\bibitem{Uniyal-2018}
R. Uniyal, H. Nandan, Ph. Jetzer,
Bending angle of light in equatorial plane of Kerr-Sen Black Hole,
Physics Letters B, 782, 185 (2018)





\bibitem{Bergliaffa-2020}
S.E.P. Bergliaffa, E.E. de Souza Filho, and R. Maier,
Strong lensing and nonminimally coupled electromagnetism,
Phys. Rev. D 101, 124038 (2020)


\bibitem{Tsukamoto-2021a}
N. Tsukamoto,
Gravitational lensing in the Simpson-Visser black-bounce spacetime in a strong deflection limit,
Phys. Rev. D 103, 024033 (2021)




\bibitem{Tsukamoto-2021b}
N. Tsukamoto,
Gravitational lensing by a photon sphere in a Reissner-Nordstr\"{o}m naked singularity spacetime in strong deflection limits,
Phys. Rev. D 104, 124016 (2021)



\bibitem{Islam-2021}
Sh. Ul Islam and S. G. Ghosh,
Strong field gravitational lensing by hairy Kerr black holes,
Phys. Rev. D 103, 124052 (2021)




\bibitem{Furtado-2021}
C. Furtado, J.R. Nascimento, A.Yu. Petrov, et al,
Strong gravitational lensing in a spacetime with topological charge within the Eddington-inspired Born-Infeld gravity,
Phys. Rev. D 103, 044047 (2021)



\bibitem{Aratore-Bozza-2021}
F. Aratore and V. Bozza,
Decoding a black hole metric from the interferometric pattern of the relativistic images of a compact source,
Journal of Cosmology and Astroparticle Physics, 10, 054 (2021)



\bibitem{Tsukamoto-2022}
N. Tsukamoto,
Retrolensing by two photon spheres of a black-bounce spacetime,
Phys. Rev. D 105, 084036 (2022)






\bibitem{Tsupko-2014}
O.Yu. Tsupko,
Unbound motion of massive particles in the Schwarzschild metric: Analytical description in case of strong deflection,
Phys. Rev. D 89, 084075 (2014)



\bibitem{Liu-2016}
X. Liu, N. Yang and J. Jia,
Gravitational lensing of massive particles in Schwarzschild gravity,
Class. Quantum Grav. 33, 175014 (2016)



\bibitem{Crisnejo-Gallo-Jusufi-2019}
G. Crisnejo, E. Gallo, and K. Jusufi,
Higher order corrections to deflection angle of massive particles and light rays in plasma media for stationary spacetimes using the Gauss-Bonnet theorem,
Phys. Rev. D, 100, 104045 (2019)




\bibitem{Wang-Liu-2020}
Y. Wang, X. Liu, N. Yang, J. Liu and J. Jia,
Escape, bound and capture geodesics in local static coordinates in Schwarzschild spacetime,
General Relativity and Gravitation, 52, 16 (2020) 







\bibitem{Bozza-Scarpetta-2007}
V. Bozza and G. Scarpetta,
Strong deflection limit of black hole gravitational lensing with arbitrary source distances,
Phys. Rev. D 76, 083008 (2007)



\bibitem{Aldi-Bozza-2017}
G.F. Aldi and V. Bozza,
Relativistic iron lines in accretion disks: the contribution of higher order images in the strong deflection limit,
Journal of Cosmology and Astroparticle Physics, 02, 033 (2017)








\bibitem{Falcke-2000}
H. Falcke, F. Melia, E. Agol,
Viewing the shadow of the black hole at the Galactic center,
Astrophys. J. Lett. 528, L13 (2000)




\bibitem{EHT-1}
K. Akiyama et al (Event Horizon Telescope Collaboration), First M87 Event Horizon Telescope Results. I. The Shadow of the Supermassive Black Hole, The Astrophysical Journal Letters, V.875, article id. L1 (2019).

\bibitem{EHT-2}
K. Akiyama et al (Event Horizon Telescope Collaboration),
First M87 Event Horizon Telescope Results. II. Array and Instrumentation, The Astrophysical Journal Letters, V.875, article id. L2 (2019).

\bibitem{EHT-3}
K. Akiyama et al (Event Horizon Telescope Collaboration),
First M87 Event Horizon Telescope Results. III. Data Processing and Calibration, The Astrophysical Journal Letters, V.875, article id. L3 (2019).

\bibitem{EHT-4}
K. Akiyama et al (Event Horizon Telescope Collaboration),
First M87 Event Horizon Telescope Results. IV. Imaging the Central Supermassive Black Hole, The Astrophysical Journal Letters, V.875, article id. L4 (2019).

\bibitem{EHT-5}
K. Akiyama et al (Event Horizon Telescope Collaboration),
First M87 Event Horizon Telescope Results. V. Physical Origin of the Asymmetric Ring, The Astrophysical Journal Letters, V.875, article id. L5 (2019).


\bibitem{EHT-6}
K. Akiyama et al (Event Horizon Telescope Collaboration),
First M87 Event Horizon Telescope Results. VI. The Shadow and Mass of the Central Black Hole, The Astrophysical Journal Letters, V.875, article id. L6 (2019).



\bibitem{Psaltis-EHT-2020}
D. Psaltis et al. (EHT Collaboration),
Gravitational Test beyond the First Post-Newtonian Order with the Shadow of the M87 Black Hole,
Phys. Rev. Lett. 125, 141104 (2020)






\bibitem{Kocherlakota-2021}
P. Kocherlakota et al. (EHT Collaboration),
Constraints on black-hole charges with the 2017 EHT observations of M87$^*$,
Phys. Rev. D 103 (2021) 104047








\bibitem{EHT-2022-1}
Event Horizon Telescope Collaboration et al,
First Sagittarius A* Event Horizon Telescope Results. I. The Shadow of the Supermassive Black Hole in the Center of the Milky Way
ApJL, 930, L12 (2022)



\bibitem{EHT-2022-2}
The Event Horizon Telescope Collaboration et al.,
First Sagittarius A* Event Horizon Telescope Results. II. EHT and Multiwavelength Observations, Data Processing, and Calibration,
ApJL, 930, L13 (2022)


\bibitem{EHT-2022-3}
The Event Horizon Telescope Collaboration et al.,
First Sagittarius A* Event Horizon Telescope Results. III. Imaging of the Galactic Center Supermassive Black Hole,
ApJL, 930, L14 (2022)


\bibitem{EHT-2022-4}
The Event Horizon Telescope Collaboration et al.,
First Sagittarius A* Event Horizon Telescope Results. IV. Variability, Morphology, and Black Hole Mass,
ApJL, 930, L15 (2022)

\bibitem{EHT-2022-5}
The Event Horizon Telescope Collaboration et al.,
First Sagittarius A* Event Horizon Telescope Results. V. Testing Astrophysical Models of the Galactic Center Black Hole,
ApJL, 930, L16 (2022)


\bibitem{EHT-2022-6}
The Event Horizon Telescope Collaboration et al.,
First Sagittarius A* Event Horizon Telescope Results. VI. Testing the Black Hole Metric,
ApJL, 930, L17 (2022)







\bibitem{Gralla-2019}
S.E. Gralla, D. E. Holz, R.M. Wald, 
Black hole shadows, photon rings, and lensing rings,
Phys. Rev. D 100, 024018 (2019)



\bibitem{Johnson-2020}
M.D. Johnson, A. Lupsasca, A. Strominger, et al, 
Universal interferometric signatures of a black hole's photon ring,
Science Advances 6, eaaz1310 (2020)




\bibitem{Gralla-Lupsasca-2020}
S.E. Gralla and A. Lupsasca,
Lensing by Kerr black holes,
Phys. Rev. D 101, 044031 (2020)





\bibitem{Gralla-Lupsasca-Marrone-2020}
S. E. Gralla, A. Lupsasca, and D. P. Marrone,
The shape of the black hole photon ring: A precise test of strong-field general relativity,
Phys. Rev. D 102, 124004 (2020)




\bibitem{Pesce-2021}
D.W. Pesce, D. C. M. Palumbo, R. Narayan, L. Blackburn, S.S. Doeleman, M. D. Johnson, et al,
Towards determining the number of observable supermassive black hole shadows,
The Astrophysical Journal, 923, 260 (2021)


\bibitem{Wielgus-2021}
M. Wielgus,
Photon rings of spherically symmetric black holes and robust tests of non-Kerr metrics,
Phys. Rev. D 104, 124058 (2021)





\bibitem{Gan-Wang-2021}
Q. Gan, P. Wang, H. Wu, and H. Yang,
Photon ring and observational appearance of a hairy black hole,
Phys. Rev. D 104, 044049 (2021)


\bibitem{Guerrero-2021}
M. Guerrero, G.J. Olmo, D. Rubiera-Garcia and D. S.-C. G\'{o}mez,
Shadows and optical appearance of black bounces illuminated by a thin accretion disk,
Journal of Cosmology and Astroparticle Physics, 08, 036 (2021)








\bibitem{Frion-2021}
E. Frion, L. Giani, T. Miranda,
Black Hole Shadow Drift and Photon Ring Frequency Drift,
The Open Journal of Astrophysics, vol. 4, issue 1, id. 9 (2021)




\bibitem{Wang-Chen-2022}
M. Wang, S. Chen, J. Jing,
Chaotic Shadows of Black Holes: A Short Review,
arXiv:2205.05855 (2022)



\bibitem{Bogush-2022}
I. Bogush, D. Gal'tsov, G. Gyulchev, K. Kobialko, P. Nedkova, T. Vetsov,
Photon surfaces, shadows and accretion disks in gravity with minimally coupled scalar field,
arXiv:2205.01919 (2022)




\bibitem{Chen-Roy-2022}
Y. Chen, R. Roy, S. Vagnozzi, L. Visinelli,
Superradiant evolution of the shadow and photon ring of Sgr A*,
arXiv:2205.06238 (2022)


\bibitem{Hou-Liu-2022}
Y. Hou, P. Liu, M. Guo, H. Yan, B. Chen,
Multi-level images around Kerr-Newman black holes,
arXiv:2203.02755 (2022)



\bibitem{Walia-2022}
R. K. Walia,
Observational Predictions of LQG Motivated Polymerized Black Holes and Constraints From Sgr A* and M87*,
arXiv:2207.02106 (2022)


\bibitem{Andrianov-2022}
A. Andrianov, S. Chernov, I. Girin, S. Likhachev, A. Lyakhovets, and Yu. Shchekinov,
Flares and their echoes can help distinguish photon rings from black holes with space-Earth very long baseline interferometry,
Phys. Rev. D 105, 063015 (2022)


\bibitem{Broderick-2022}
A.E. Broderick, P. Tiede, D.W. Pesce, R. Gold,
Measuring Spin from Relative Photon Ring Sizes,
The Astrophysical Journal, 927, 6 (2022) 



\bibitem{Paugnat-Lupsasca-2022}
H. Paugnat, A. Lupsasca, F. Vincent, M. Wielgus,
Photon ring test of the Kerr hypothesis: variation in the ring shape,
arXiv: 2206.02781 (2022)



\bibitem{Vincent-Gralla-2022}
F. H. Vincent, S. E. Gralla, A. Lupsasca, M. Wielgus,
Images and photon ring signatures of thick disks around black holes,
arXiv:2206.12066 (2022)


\bibitem{Ayzenberg-2022}
D. Ayzenberg,
Testing gravity with black hole shadow subrings,
Classical and Quantum Gravity, Volume 39, Issue 10, id.105009 (2022)





\bibitem{Broderick-Loeb-2005}
A. E. Broderick, A. Loeb,
Imaging bright-spots in the accretion flow near the black hole horizon of Sgr A*,
Mon. Not. Roy. Astron. Soc. 363 (2005) 353


\bibitem{Moscibrodzka-2009}
M. Mo\'{s}cibrodzka, Ch.F. Gammie, J. C. Dolence, H. Shiokawa, Po Kin Leung,
Radiative models of Sgr A* from GRMHD simulations,
Astrophys. J. 706 (2009) 497



\bibitem{Dexter-2009}
J. Dexter, E. Agol, P. Chris Fragile,
Millimeter flares and VLBI visibilities from relativistic simulations of magnetized accretion onto the Galactic center black hole,
Astrophys. J. Lett. 703  (2009) L142




\bibitem{Broderick-Johannsen-2014}
A.E. Broderick, T. Johannsen, A. Loeb, D. Psaltis,
Testing the no-hair theorem with Event Horizon Telescope observations of Sagittarius A*,
Astrophys. J. 784 (2014) 7


\bibitem{James-2015}
O. James, E. von Tunzelmann, P. Franklin, K. S. Thorne, 
Gravitational lensing by spinning black holes in astrophysics, and in the movie \emph{Interstellar},
Class. Quantum Grav. 32, 065001 (2015)







\bibitem{Narayan-2019}
R. Narayan, M.D. Johnson, C.F. Gammie, 
The shadow of a spherically accreting black hole
Astrophys. J. Lett. 885, L33 (2019) 




\bibitem{Dokuchaev-2020}
V.I. Dokuchaev, N.O. Nazarova,
Silhouettes of invisible black holes,
Physics-Uspekhi 63 (2020)  583




\bibitem{Bronzwaer-Falcke-2021}
Th. Bronzwaer and H. Falcke,
The Nature of Black Hole Shadows,
The Astrophysical Journal, Volume 920, Issue 2, id.155, 12 pp. (2021)




\bibitem{Andrianov-2021}
A. S. Andrianov, A. M. Baryshev, H. Falcke, et al,
Simulations of M87 and Sgr A* imaging with the Millimetron Space Observatory on near-Earth orbits,
Monthly Notices of the Royal Astronomical Society 500 (2021) 4866



\bibitem{Vincent-2021}
F. H. Vincent, M. Wielgus, M. A. Abramowicz, E. Gourgoulhon, et al,
Geometric modeling of M87* as a Kerr black hole or a non-Kerr compact object,
Astronomy and Astrophysics, Volume 646, id.A37 (2021)




\bibitem{Bronzwaer-Davelaar-2021}
Th. Bronzwaer, J. Davelaar, Z. Younsi et al, 
Visibility of black hole shadows in low-luminosity AGN,
Mon. Not. Roy. Astron. Soc. 501 (2021) 4722





\bibitem{Bardeen-1973}
J.~M. Bardeen, Timelike and null geodesics in the Kerr metric, in {\em Black Holes}, eds. C. DeWitt and B. DeWitt (Gordon and Breach, New York, 1973), p. 215.




\bibitem{Chandra-1983}
S. Chandrasekhar,
\textit{The Mathematical Theory of Black Holes} (Oxford University Press, Oxford, 1983).




\bibitem{Cunha-Herdeiro-2018}
P.V.P. Cunha and C.A.R. Herdeiro,
Shadows and strong gravitational lensing: a brief review,
Gen Relativ Gravit 50, 42 (2018). 


\bibitem{Perlick-Tsupko-2022}
V. Perlick and O.Yu. Tsupko,
Calculating black hole shadows: review of analytical studies,
Physics Reports, 947, 1 (2022)




\bibitem{Gren-Perlick-2014}
A. Grenzebach, V. Perlick and C. L{\"a}mmerzahl,
Photon regions and shadows of Kerr-Newman-NUT black holes with a cosmological constant,
Phys. Rev. D 89, 124004 (2014)


\bibitem{Gren-Perlick-2015}
A. Grenzebach, V. Perlick and C. L{\"a}mmerzahl,
Photon regions and shadows of accelerated black holes,
International Journal of Modern Physics D, 24, 1542024 (2015)








\bibitem{Young-1976}
P. J. Young,
Capture of particles from plunge orbits by a black hole,
Phys. Rev. D 14 (1976) 3281



\bibitem{Dymnikova-1986}
I. G. Dymnikova,
Motion of particles and photons in the gravitational field of a rotating body (In memory of Vladimir Afanas'evich Ruban),
Soviet Physics Uspekhi, 29, 215 (1986).



\bibitem{Vries-2000}
A. de Vries,
The apparent shape of a rotating charged black hole, closed photon orbits and the bifurcation set $A_4$,
Class. Quantum Grav. 17 (2000)  123










\bibitem{Takahashi-2004}
R. Takahashi,
Shapes and Positions of Black Hole Shadows in Accretion Disks and Spin Parameters of Black Holes,
Astrophys. J., 611, 996 (2004).




\bibitem{Zakharov-Paolis-2005-New-Astronomy}
A.F. Zakharov, A.A. Nucita, F. DePaolis, G. Ingrosso,
Measuring the black hole parameters in the galactic center with RADIOASTRON,
New Astronomy, 10, 479 (2005).


\bibitem{Hioki-Maeda-2009}
K. Hioki and K. Maeda,
Measurement of the Kerr spin parameter by observation of a compact object's shadow,
Phys. Rev. D 80 (2009) 024042 






\bibitem{Johannsen-2010}
T. Johannsen and D. Psaltis,
Testing the No-hair Theorem with Observations in the Electromagnetic Spectrum. II. Black Hole Images,
The Astrophysical Journal, 718, 446 (2010)



\bibitem{Frolov-Zelnikov-2011}
V. P. Frolov, A. Zelnikov, 
Introduction to Black Hole Physics,
Oxford Univ. Press, New York (2011)


\bibitem{Johannsen-2013}
T. Johannsen,
Photon Rings around Kerr and Kerr-like Black Holes,
The Astrophysical Journal, 777, 170 (2013)


\bibitem{Li-Bambi-2014}
Z. Li, C. Bambi,
Measuring the Kerr spin parameter of regular black holes from their shadow,
J. Cosm. Astropart. Phys., 01, 041 (2014)





\bibitem{Tsupko-2017}
O. Yu. Tsupko,
Analytical calculation of black hole spin using deformation of the shadow,
Phys. Rev. D 95, 104058 (2017)


\bibitem{Gralla-Lups-2018}
S.E. Gralla, A. Lupsasca and A. Strominger,
Observational signature of high spin at the Event Horizon Telescope,
Mon. Not. Roy. Astron. Soc. 475 (2018) 3829



\bibitem{Wei-2019-Rapid}
Sh.-W. Wei, Y.-X. Liu, and R.B. Mann,
Intrinsic curvature and topology of shadows in Kerr spacetime,
Phys. Rev. D, 99, 041303(R) (2019)









\bibitem{Farah-2020}
J. R. Farah, D. W. Pesce, M. D. Johnson, and L. Blackburn,
On the Approximation of the Black Hole Shadow with a Simple Polar Curve,
The Astrophysical Journal, 900, 77 (2020)


\bibitem{Gralla-Lups-2020c}
S.E. Gralla, A. Lupsasca,
Observable shape of black hole photon rings,
Phys. Rev. D 102 (2020)  124003




\bibitem{Perlick-Tsupko-BK-2015}
V. Perlick, O. Yu. Tsupko, G. S. Bisnovatyi-Kogan,
Influence of a plasma on the shadow of a spherically symmetric black hole,
Phys. Rev. D 92, 104031 (2015)



\bibitem{Cunha-2015}
P.V.P. Cunha, C.A.R. Herdeiro, E. Radu, H.F. R{\'u}narsson,
Shadows of Kerr Black Holes with Scalar Hair,
Phys. Rev. Lett. 115, 211102 (2015)



\bibitem{Abdujabbarov-2015}
A. A. Abdujabbarov, L. Rezzolla, B. J. Ahmedov,
A coordinate-independent characterization of a black hole shadow, Monthly Notices of the Royal Astronomical Society, 454, 2423 (2015)





\bibitem{Shipley-Dolan-2016}
J. O. Shipley and S. R. Dolan,
Binary black hole shadows, chaotic scattering and the Cantor set,
Class. Quantum Grav. 33, 175001 (2016)





\bibitem{Perlick-Tsupko-2017}
V. Perlick, O. Yu. Tsupko,
Light propagation in a plasma on Kerr space-time: Separation of the Hamilton-Jacobi equation and calculation of the shadow,
Phys. Rev. D 95, 104003 (2017) 



\bibitem{Yan-2018}
H. Yan,
Influence of a plasma on the observational signature of a high-spin Kerr black hole,
Phys. Rev. D 99, 084050 (2019)



\bibitem{Mars-2018}
M. Mars, C.F. Paganini and M.A. Oancea,
The fingerprints of black holes - shadows and their degeneracies,
Classical and Quantum Gravity, 35, 025005 (2018) 


\bibitem{Yunes-2018}
D. Ayzenberg and N. Yunes,
Black hole shadow as a test of general relativity: quadratic gravity,
Classical and Quantum Gravity, 35, 235002 (2018).



\bibitem{Cunha-2018}
P. V. P. Cunha, C. A. R. Herdeiro, and M. J. Rodriguez,
Does the black hole shadow probe the event horizon geometry?
Phys. Rev. D 97, 084020 (2018)





\bibitem{Perlick-Tsupko-BK-2018}
V. Perlick, O. Yu. Tsupko, and G. S. Bisnovatyi-Kogan,
Black hole shadow in an expanding universe with a cosmological constant,
Physical Review D, 97, 104062 (2018)




\bibitem{BK-Tsupko-2018}
G. S. Bisnovatyi-Kogan and O. Yu. Tsupko,
Shadow of a black hole at cosmological distances,
Physical Review D, 98, 084020 (2018)


\bibitem{Abdikamalov-2019}
A. B. Abdikamalov, A. A. Abdujabbarov, D. Ayzenberg, D. Malafarina, C. Bambi, and B. Ahmedov,
Black hole mimicker hiding in the shadow: Optical properties of the $\gamma$ metric,
Phys. Rev. D, 100, 024014 (2019)








\bibitem{Tsupko-Fan-BK-2020}
O.Yu. Tsupko, Z. Fan, and G.S. Bisnovatyi-Kogan,
Black hole shadow as a \textit{standard ruler} in cosmology,
Classical and Quantum Gravity, 37, 065016 (2020); arXiv:1905.10509


\bibitem{Qi-Zhang-2020}
J.-Zh. Qi, X. Zhang,
A new cosmological probe using super-massive black hole shadows,
Chinese Phys. C 44 (2020) 055101









\bibitem{Vagnozzi-2020}
S. Vagnozzi, C. Bambi, L. Visinelli,
Concerns regarding the use of black hole shadows as standard rulers,
Class. Quantum Grav., 37, 087001 (2020)





\bibitem{Neves-2020}
J.C.S. Neves,
Constraining the tidal charge of brane black holes using their shadows,
The European Physical Journal C, 80, 717 (2020)




\bibitem{Li-Guo-2020}
P.-C. Li, M. Guo, and B. Chen,
Shadow of a spinning black hole in an expanding universe,
Phys. Rev. D 101, 084041 (2020)




\bibitem{Chang-Zhu-2020}
Zh. Chang and Q.-H. Zhu,
Black hole shadow in the view of freely falling observers,
Journal of Cosmology and Astroparticle Physics, 06,
055 (2020)


\bibitem{Tsupko-BK-2020-IJMPD}
O.Yu. Tsupko and G.S. Bisnovatyi-Kogan,
First analytical calculation of black hole shadow in McVittie metric,
International Journal of Modern Physics D, 29, 2050062 (2020)





\bibitem{Kumar-Ghosh-2020}
R. Kumar and S.G. Ghosh,
Black Hole Parameter Estimation from Its Shadow,
The Astrophysical Journal, 892, 78 (2020)



\bibitem{Cunha-2020}
P. V. P. Cunha and C. A. R. Herdeiro,
Stationary Black Holes and Light Rings,
Phys. Rev. Lett. 124, 181101 (2020)






\bibitem{Lima-2021}
H. C. D. Lima Junior, L. C. B. Crispino, P. V. P. Cunha, and C. A. R. Herdeiro,
Can different black holes cast the same shadow?
Phys. Rev. D 103, 084040 (2021)


\bibitem{Devi-2021}
S. Devi, S. Chakrabarti, B. R. Majhi,
Shadow of quantum extended Kruskal black hole and its super-radiance property,
arXiv:2105.11847 (2021)




\bibitem{Pantig-2021}
R. C. Pantig, P. K. Yu, E. T. Rodulfo, Ali \"{O}vg\"{u}n,
Shadow and weak deflection angle of extended uncertainty principle black hole surrounded with dark matter,
arXiv:2104.04304 (2021)









\bibitem{Anacleto-2021}
M. A. Anacleto, J. A. V. Campos, F. A. Brito, E. Passos,
Quasinormal modes and shadow of a Schwarzschild black hole with GUP,
arXiv: 2108.04998 (2021)




\bibitem{Khodadi-2021}
M. Khodadi, G. Lambiase, D. F. Mota,
No-Hair Theorem in the Wake of Event Horizon Telescope,
arXiv: 2107.00834 (2021)







\bibitem{Tsupko-2021}
O.Yu. Tsupko,
Deflection of light rays by a spherically symmetric black hole in a dispersive medium,
Phys. Rev. D 103, 104019 (2021)


\bibitem{Cardoso-2021}
V. Cardoso, F. Duque, and A. Foschi,
Light ring and the appearance of matter accreted by black holes,
Phys. Rev. D 103, 104044 (2021)










\bibitem{Wang-2021}
J. Wang,
Multiple rings in the shadow of extremely compact objects,
International Journal of Modern Physics D, 30, 2150112 (2021)


\bibitem{Eichhorn-2021}
A. Eichhorn and A. Held,
Image features of spinning regular black holes based on a locality principle,
The European Physical Journal C, 81, 933 (2021)




\bibitem{Ozel-2021}
F. Ozel, D. Psaltis, Z. Younsi,
Black Hole Images as Tests of General Relativity: Effects of Plasma Physics,
arXiv:2111.01123 (2021)


\bibitem{Roy-2021}
R. Roy, S. Vagnozzi, L. Visinelli,
Superradiance evolution of black hole shadows revisited,
arXiv:2112.06932 (2021)




\bibitem{Ghosh-Sarkar-2021}
R. Ghosh and S. Sarkar,
Light rings of stationary spacetimes,
Phys. Rev. D 104, 044019 (2021)



\bibitem{Uniyal-Pantig-2022}
A. Uniyal, R. C. Pantig and A. \"{O}vg\"{u}n,
Probing a nonlinear electrodynamics black hole with thin accretion disk, shadow and deflection angle with M87* and Sgr A* from EHT,
eprint arXiv: 2205.11072 (2022)



\bibitem{Vagnozzi-review-2022}
S. Vagnozzi, R. Roy, Y.-D. Tsai, et al,
Horizon-scale tests of gravity theories and fundamental physics from the Event Horizon Telescope image of Sagittarius A*,
eprint arXiv: 2205.07787 (2022)





\bibitem{Luminet-1979}
J.-P. Luminet, 
Image of a spherical black hole with thin accretion disk,
Astron. Astrophys. 75, 228 (1979)



\bibitem{Cunningham-1973}
C.T. Cunningham, J.M. Bardeen,
The optical appearance of a star orbiting an extreme Kerr black hole
Astrophys. J. 183 (1973) 237 


\bibitem{Hadar-Johnson-2021}
S. Hadar, M.D. Johnson, A. Lupsasca, and G.N. Wong,
Photon ring autocorrelations,
Phys. Rev. D 103, 104038 (2021)




\end{thebibliography}

\end{document}